\documentclass[twocolumn]{aastex631}

\usepackage{subfigure} 
\usepackage{amssymb}
\usepackage[para]{threeparttable}
\usepackage{hyperref}

\received{November 16, 2023}
\revised{January 11, 2024}
\accepted{January 8, 2024}

\shorttitle{Stellar Loci. VII. Photometric Metallicities Based on GALEX and Gaia}
\shortauthors{Lu et al.}

\begin{document}

\title{Stellar Loci. VII. Photometric Metallicities of 5 Million FGK Stars Based on GALEX GR6+7 AIS and Gaia EDR3}

\author[0009-0009-0678-2537]{Xue Lu}
\affiliation{Institute for Frontiers in Astronomy and Astrophysics, Beijing Normal University,  Beijing 102206, China}
\affiliation{Department of Astronomy, Beijing Normal University No.19, Xinjiekouwai St, Haidian District, Beijing, 100875, China}

\author[0000-0003-2471-2363]{Haibo Yuan}
\affiliation{Institute for Frontiers in Astronomy and Astrophysics, Beijing Normal University,  Beijing 102206, China}
\affiliation{Department of Astronomy, Beijing Normal University No.19, Xinjiekouwai St, Haidian District, Beijing, 100875, China}

\author[0000-0003-3535-504X]{Shuai Xu}
\affiliation{Institute for Frontiers in Astronomy and Astrophysics, Beijing Normal University,  Beijing 102206, China}
\affiliation{Department of Astronomy, Beijing Normal University No.19, Xinjiekouwai St, Haidian District, Beijing, 100875, China}

\author[0000-0003-1863-1268]{Ruoyi Zhang}
\affiliation{Institute for Frontiers in Astronomy and Astrophysics, Beijing Normal University,  Beijing 102206, China}
\affiliation{Department of Astronomy, Beijing Normal University No.19, Xinjiekouwai St, Haidian District, Beijing, 100875, China}

\author[0000-0001-8424-1079]{Kai Xiao}
\affiliation{Institute for Frontiers in Astronomy and Astrophysics, Beijing Normal University,  Beijing 102206, China}
\affiliation{Department of Astronomy, Beijing Normal University No.19, Xinjiekouwai St, Haidian District, Beijing, 100875, China}
\affiliation{School of Astronomy and Space Science, University of Chinese Academy of Sciences, Beijing 100049, China}

\author[0000-0003-3250-2876]{Yang Huang}
\affiliation{School of Astronomy and Space Science, University of Chinese Academy of Sciences, Beijing 100049, China}

\author[0000-0003-4573-6233]{Timothy C. Beers}
\affiliation{Department of Physics and Astronomy, University of Notre Dame, Notre Dame, IN 46556, USA}
\affiliation{Joint Institute for Nuclear Astrophysics -- Center for the Evolution of the Elements (JINA-CEE), USA}

\author[0000-0002-2453-0853]{Jihye Hong}
\affiliation{Department of Physics and Astronomy, University of Notre Dame, Notre Dame, IN 46556, USA}
\affiliation{Joint Institute for Nuclear Astrophysics -- Center for the Evolution of the Elements (JINA-CEE), USA}

\correspondingauthor{Haibo Yuan}
\email{yuanhb@bnu.edu.cn}

\begin{abstract}
We combine photometric data from GALEX GR6+7 AIS and Gaia EDR3 with stellar parameters from the SAGA and PASTEL catalogs to construct high-quality training samples for dwarfs ($\rm 0.4< BP-RP<1.6$) and giants ($\rm 0.6< BP-RP <1.6$). We apply careful reddening corrections using empirical temperature- and extinction-dependent extinction coefficients. Using the two samples, we establish a relationship between stellar loci (NUV$-$BP vs. BP$-$RP colors), metallicity, and $\rm M_G$. For a given BP$-$RP color, a 1\,dex change in [Fe/H] corresponds to an approximately 1 magnitude change in NUV$-$BP color for Solar-type stars. These relationships are employed to estimate metallicities based on NUV$-$BP, BP$-$RP, and $\rm M_G$. Thanks to the strong metallicity dependence in the GALEX NUV-band, our models enable a typical photometric-metallicity precision of approximately $\sigma_{\rm [Fe/H]}$ = 0.11\,dex for dwarfs and $\sigma_{\rm [Fe/H]}$ = 0.17\,dex for giants, with an effective metallicity range extending down to [Fe/H] $= -3.0$ for dwarfs and [Fe/H] $= -4.0$ for giants. We also find that the NUV-band based photometric-metallicity estimate is not as strongly affected by carbon enhancement as previous photometric techniques.
With the GALEX and Gaia data, we have estimated metallicities for about 5 million stars across almost the entire sky, including approximately 4.5 million dwarfs and 0.5 million giants. This work demonstrates the potential of the NUV-band for estimating photometric metallicities, and sets the groundwork for utilizing the NUV data from space telescopes such as the upcoming Chinese Space Station Telescope. 
\end{abstract}

\keywords{metallicity; stellar fundamental parameters; astronomy data analysis}

\section{Introduction} \label{sec:intro}

Metallicity (metal abundance; generally parametrized by [Fe/H]), is one of the basic stellar parameters, and plays an important role in studying the formation and evolution of not only stars and stellar populations, but also of galaxies such as the Milky Way. Analyzing the
stellar-metallicity distribution within the Milky Way provides valuable insights into its origin and evolution through chemical and kinematic studies (e.g., \citealt{Beers_2005metelpoor,Matteucci_2021}). 

There are two main methods to estimate stellar metallicity for large samples of stars utilized to date: spectroscopic and photometric. Spectroscopy has long been the primary approach, and significant progress has been made with large-scale spectroscopic surveys over the past two decades. Notable surveys include the Sloan Digital Sky Survey and the Sloan Extension for Galactic Understanding and Evolution (SDSS/SEGUE; \citealt{SDSS2000,Yanny2009,Rockosi2022}), the Radial Velocity Experiment (RAVE; \citealt{RAVE2006}), the Large Sky Area Multi-Object Fiber Spectroscopic Telescope (LAMOST; \citealt{Deng2012,Liu2014}), the Gaia-ESO Public Spectroscopic Survey (\citealt{Gilmore_2022,Randich_2022}), the Apache Point Observatory Galactic Evolution Experiment (APOGEE; \citealt{Majewski2017}), and the GALactic Archaeology with HERMES (GALAH; \citealt{GALAH2018}). Moderate ($R \sim 7,500$)- to high ($R \gtrsim 20,000)$-resolution spectroscopy can achieve high-precision metallicity estimates, typically around or below 0.1\,dex, especially in cases of high spectral signal-to-noise ratios. However, there are also disadvantages to spectroscopy. For example, spectroscopic surveys are more time-consuming and require complex data analysis compared to photometric surveys. Consequently, the number of observed sources collected from the aforementioned surveys represents only a tiny fraction of the estimated total number of stars in the Milky Way. The resulting limited samples, coupled with typically complex selection functions for the targets, poses challenges to obtaining a more complete and unbiased set of metallicity estimates for stars in the Milky Way.
   
Stellar metallicity can also be obtained by analyzing photometric data, after proper calibration from spectroscopy. Original efforts can be traced back to \cite{Schwarzschild_1955}, but steady progress in this approach has been made over the intervening decades due to improvements in technology and the use of optimized filters. Stellar colors are not a pure function of the effective temperature; other stellar parameters also have an impact, including the stellar metallicity. Though it has a smaller effect on the stellar color than temperature, the metallicity of a star can still be inferred photometrically. The dependence of stellar loci on metallicity is more pronounced in bluer colors, as metallic absorption lines in the optical region are primarily concentrated in this wavelength range. 

A typical procedure for obtaining photometric-metallicity estimates involves several steps. First, cross-matches of photometric catalogs of stars with available spectroscopic-metallicity estimates are performed. Data quality cuts and careful reddening corrections are then applied for the stars in common. Subsequently, various techniques such as metallicity-dependent stellar loci fitting (e.g., \citealt{yuan2015metal}, Paper I: the 1st paper of this series), principal component analysis (e.g., \citealt{Casagrande2019}), stellar isochrone fitting (e.g., \citealt{An&Beers2020}), neural networks (e.g., \citealt{Whitten2021,Yang2022}), and others are employed to establish the relationship between stellar colors and metallicities. This relationship is then further validated and tested using other independent sets of data. Finally, the established relationship is applied to all photometric data that satisfy certain conditions, enabling the determination of photometric metallicities for those sources. Furthermore, it is important to recognize that the relationship between stellar colors and metallicities varies for stars of different luminosity classes.
    
Among these methods, the stellar loci fitting method stands out as one of the earliest, and indeed, it has been widely employed
(e.g., \citealt{Ivezic2008,yuan2015metal,Starkenburg2017,Huang2019,huang2021}). 
\cite{Ivezic2008} used the SDSS colors $u - g$ and $g - r$ to provide photometric metallicities for 2 million dwarfs with a precision of 0.15\,dex, down to a metallicity limit of [Fe/H] $\sim -2.0$. In a later study, \citeauthor{yuan2015metal} (\citeyear{yuan2015metal}, Paper I) investigated the metallicity dependence of the stellar loci from SDSS and their intrinsic widths after considering effect of metallicity and photometric error, by using re-calibrated photometric data from SDSS stripe 82 (\citealt{yuan2015recalibrated}). It was found that a 1\,dex decrease in metallicity results in a decrease of about 0.20/0.02 mag in the $u - g/g - r$ colors and increase of about 0.02/0.02 mag in the $r - i/i - z$ colors, respectively. Additionally, \citeauthor{yuan2015FGK} (\citeyear{yuan2015FGK}, Paper III) estimated photometric metallicities for half a million FGK stars from SDSS stripe 82, with a typical precision of 0.1\,dex. by simultaneously fitting de-reddened $u - g$, $g - r$, $r - i$, and $i - z$ colors from SDSS to those predicted for metallicity-dependent stellar loci. After that, \citeauthor{zhang2021} (\citeyear{zhang2021}, Paper IV) studied metallicity-dependent SDSS stellar color loci for red giant stars, and measured stellar metallicities in Stripe 82 using the same technique, achieving a typical precision of 0.2\,dex. Notably, they observed systematic differences between the metallicity-dependent stellar loci of red giants and main-sequence stars.

\cite{Huang2019,huang2022,Huang2023} apply a similar technique to the photometric data of the SkyMapper Southern Survey (SMSS; \citealt{SkyMapper_DR2}) DR2 and the Stellar Abundances and Galactic Evolution Survey (SAGES; \citealt{SAGES_DR1}) DR1, whose optimally designed narrow/medium-band u,v filters enabled photometric-metallicity estimates down to [Fe/H] $\sim -3.5$ for a total of some 50 million stars.
 
Broad-band photometry is generally believed to be unsuitable for photometric-metallicity estimation, due to its very weak sensitivity to changes in stellar metallicity. However, by combining data from Gaia EDR3 (\citealt{Brown2021}) and LAMOST DR7 (\citealt{Luo2015}), and using the stellar loci fitting technique, \citeauthor{Xu2022gaia-metal} (\citeyear{Xu2022gaia-metal}, Paper V) obtained metallicity estimates for about 27 million stars (including over 20 million dwarfs and 6 million giants) with $\rm 10<G<16$ across almost the entire sky. They found that, for a given $\rm (BP-RP)$ color, a 1\,dex change in [Fe/H] results in a 5 mmag change in the $\rm (BP-G)$ color for Solar-type stars. Despite the very weak sensitivity, the exquisite data quality of the Gaia colors, together with careful reddening corrections and color calibrations, enable a typical precision of approximately $\delta[\rm Fe/H] = 0.2$\,dex. Utilizing an extensive dataset of photometric metallicities and considering additional criteria, \citeauthor{Xu2022metal-poor} (\citeyear{Xu2022metal-poor}, Paper VI) compile an updated catalog of the Best and Brightest metal-poor stars. This catalog offers valuable targets for subsequent high-resolution spectroscopic observations.
Note that, with the data release of Gaia DR3, particularly the XP spectra, a number of massive, all-sky catalogs of metallicities have been published (\citealt{Andrae_R_2023, Martin2023, Recio-Blanco_2023, Zhang_xiangyu_2023}). 
 
Based on previous work on photometric metallicity, bluer and narrower bands are more sensitive to metallicity; the availability of high-precision photometric data is also highly advantageous for accurate and precise estimates of photometric metallicity. However, thus far, the capability of the near-ultraviolet (NUV) band (with a bandwidth on the order of 1000\,{\AA}) data for estimation of photometric metallicities has not yet been fully explored. In this study, we aim to develop the utility of NUV data from the Galaxy Evolution Explorer (GALEX; \citealt{galex}) for this purpose.
    
The paper is organized as follows. In Sections \ref{sec:data} and \ref{sec:method}, we introduce the data and the method used in this work. The results and a discussion are described in Section \ref{sec:resu-disc}. The final sample is described in Section \ref{sec:final}. Section \ref{sec:summary} presents a summary.

\section{Data} 
\label{sec:data}

\subsection{GALEX GR6+7 AIS} \label{subsec:galex}
GALEX was a NASA Explorer Mission, launched on April 28, 2003, with the primary objective of conducting the first space-based sky survey in both the near and far ultraviolet. 
GALEX surveys include an all-sky imaging survey, and medium and deep imaging surveys covering specific areas, as well as spectroscopic surveys utilizing grism technology.
GALEX simultaneously performs sky surveys in two bands through use of a dichroic beam splitter: far-UV (FUV; $1344-1786 \rm \AA$) and near-UV (NUV; $1771-2831 \rm \AA$).
The field-of-view of GALEX is 1.28$^\circ$/1.24$^\circ$, with a spatial resolution of $4.2^{\prime \prime}/5.3^{\prime \prime}$ for the FUV/NUV, respectively (\citealt{Morrissey_2007}). The GALEX GR6+7 All-Sky Imaging Survey (AIS) database (\citealt{galex_gr6/7}) is a valuable catalog derived from the mission, containing 214,449,551 source measurements. The photometric precision of the NUV-band varies from 0.0025 to 0.5 mag. Most stars brighter than 20th magnitude in the NUV have a photometric precision better than 0.15 mag.
    
\subsection{Gaia Early Data Release 3} \label{subsec:gaia}
The Gaia Early Data Release 3 (EDR3; \citealt{Brown2021}) provides unprecedented photometric data with mili-magnitude precision for over one billion stars in the $G$, $G_{\rm BP}$ and $G_{\rm RP}$ bands\footnote {For simplicity of notation, we refer to the  $G_{\rm BP}$ and $G_{\rm RP}$ magnitudes as `BP' and `RP' in the remainder of this paper.}. It also provides trigonometric parallaxes and proper motions for nearly 1.5 billion sources. This vast and high-quality dataset serves as an excellent resource for estimating photometric metallicities for an enormous number of stars. Given that the Gaia
BP and RP magnitudes are derived from aperture photometry, the `$\rm phot\_bp\_rp\_excess\_factor$' parameter, defined as $\rm (I_{BP} + I_{RP})/I_G$, serves as an indicator of the extent to which background and contamination affect the accuracy of BP/RP magnitudes.

\begin{figure}[htbp]
\centering
\includegraphics[width=1.0\linewidth]{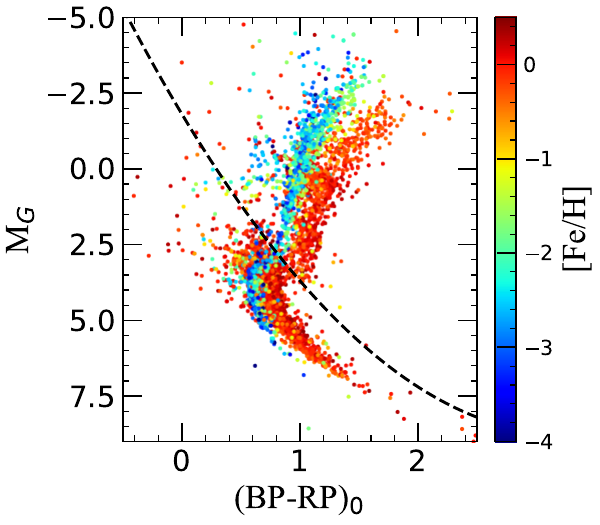}
\caption{H-R diagram of the training samples, color-coded by [Fe/H]. The dwarfs and giants are separated by the black-dashed line (\citealt{Xu2022gaia-metal}): $\rm M_G = -{(BP-RP)}^2 + 6.5\times (BP-RP) - 1.8$.}
\label{fig:HR-diagram}
\end{figure}

\subsection{Construction of Training Samples} \label{subsec:sample}
We obtain spectroscopic data for the training samples by combining metallicity information  from the Stellar Abundances for the Galactic Archeology (SAGA; \citealt{saga2008,saga2011,saga2013,saga2017}) and the PASTEL (\citealt{pastel2010,pastel2016}) catalogs. After removing old and duplicate sources (references prior to 1986 or those without [Fe/H] values are removed, and duplicate references are averaged for their [Fe/H] 
values), the number of stars in the combined dataset is 25,667. After performing a cross-match with the GALEX GR6+7 AIS and Gaia EDR3 catalogs, we obtain NUV/BP/RP-band information for 4,026 sources, which are used to construct our final training samples. There are 1,680 and 19,961 sources that are removed when cross-matching with Gaia and GALEX, respectively. Note that the radius for all cross-matching is set to be 
1\arcsec\ in this paper. Considering that the stellar loci in the NUV-band are highly sensitive to metallicity, and the sensitivity differs between giants and dwarfs (\citealt{zhang2021}), we divide the selected sources into these two categories using an empirical cut (\citealt{Xu2022gaia-metal}), as shown in Figure \ref{fig:HR-diagram}. The cuts for the training samples are as follows:

\begin{enumerate}
\item $\rm 0.4<(BP-RP)_0<1.6$ for dwarfs and $\rm 0.6<(BP-RP)_0<1.6$ for giants
\item $-3.5<[\rm Fe/H]<+0.5$ for dwarfs (considering the metallicity dependency is too weak to estimate metallicity for dwarf stars with $\rm [Fe/H]<-3.5$ using the GALEX NUV data) 
and $\rm -5<[Fe/H]<+0.5$ for giants
\item $\rm NUV>15$ to avoid saturation (\citealt{Morrissey_2007(galex_saturation)}) and $\rm error_{NUV}<0.1$ to ensure the NUV data quality
\item $\rm phot\_bp\_rp\_excess\_factor < 0.09 \times (BP-RP)+1.15$ to ensure the BP/RP data quality (the same as in \citealt{Xu2022gaia-metal})
\item $\rm E(B-V)<0.15$. The $\rm E(B-V)$ used here is from the \citeauthor{SFD98} (\citeyear{SFD98}, hereafter SFD98) dust reddening map. Note that the 14\% systematically 
over-estimated (\citealt{Schlafly_2010,Schlafly_2011,Yuan_2013}) values have been corrected in the empirical reddening coefficients described in Section \ref{sec:method}. The SFD98 map is used throughout this work, as it is one of the best choices for reddening correction of  stars at middle and high Galactic latitudes (\citealt{Sun_2022}). 

\end{enumerate}

\begin{figure}[htbp]
\centering
\includegraphics[width=1.0\linewidth]{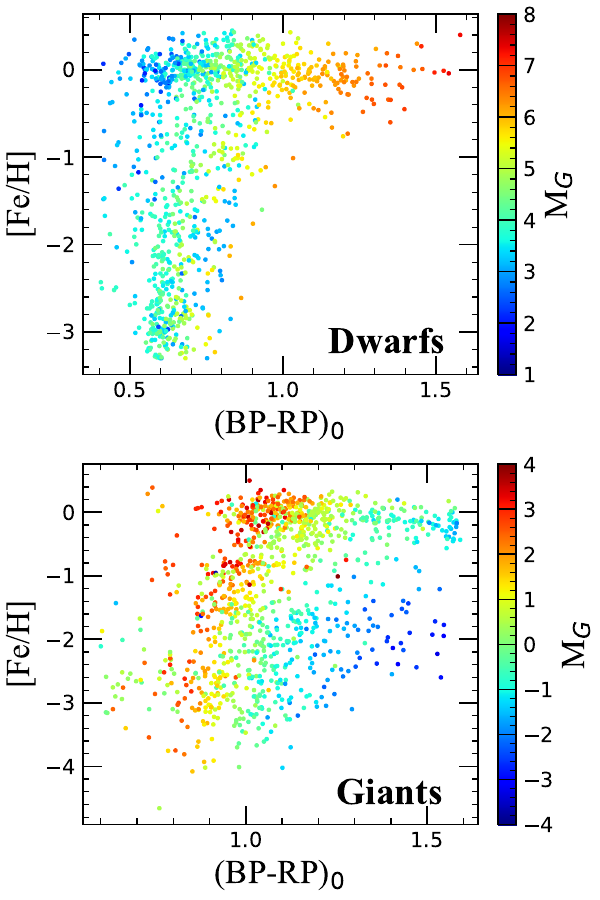}
\caption{Distributions of the dwarf (top panel) and giant (bottom panel) training samples after cutting in the [Fe/H] vs. $\rm (BP-RP)_0$ plane, color-coded by $\rm M_G$.}
\label{fig:training_sample}
\end{figure}

After applying these cuts, the dwarf and giant samples have a very limited number of sources with $-2.5 <[\rm Fe/H]< -0.3$. We thus add 200 high-quality dwarfs and giants from LAMOST with $-2.5 <[\rm Fe/H]< -0.3$, $\rm SNRg\, (LAMOST)>40$, 
$\rm error_{NUV} < $ 0.05, E(B-V) $<0.05$, and $\rm NUV>15$ into the dwarf and giant training samples.
There are small metallicity offsets between the LAMOST and the previous training samples: $-$0.07\,dex for dwarfs and $-$0.06\,dex for giants. Finally, a total of 1073 and 1107 sources are selected as our dwarf and giant training samples, respectively. Their distributions in the [Fe/H] vs. (BP$-$RP)$_0$ plane are shown in Figure \ref{fig:training_sample}. 


\section{Method} \label{sec:method}
\subsection{Reddening Corrections} \label{subsec:redden}
Due to the ultra-broad Gaia passbands, it is important to perform careful reddening corrections. \citet{zhang2023} consider the dependence of reddening coefficients on $\rm T_{ eff}$ and $\rm E(B-V)$, providing a better description of empirical reddening coefficients than simply single-valued coefficients. 
According to their results, the reddening coefficients exhibit a stronger dependence on $\rm T_{eff}$ for broader filters (such as the Gaia passbands) and bluer filters (such as the GALEX filters). In this work, we utilize the python package provided by \citet{zhang2023} to obtain $\rm T_{eff}$- and $\rm E(B-V)$-dependent reddening coefficients: $\rm R(T_{eff},E(B-V))$ for the BP$-$RP and NUV$-$BP colors. The intrinsic colors of stars are estimated as follows:
\[
color_0=color_{obs}-R_{color}(T_{\rm eff}, E(B-V))\times E(B-V).
\]
    
\begin{figure}[htbp]
\centering
\includegraphics[width=1.0\linewidth]{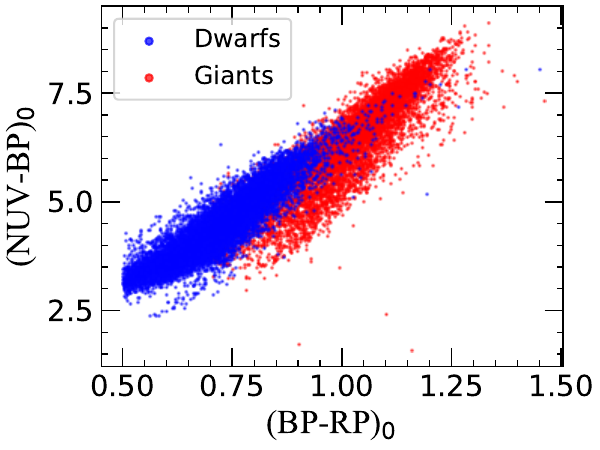}
\caption{The differences in $\rm BP-RP$ vs. $\rm NUV-BP$ color loci between dwarfs (blue dots) and giants (red dots).}
\label{fig:loci-diff}
\end{figure}

\begin{figure}[htbp]
\centering
\includegraphics[width=1.0\linewidth]{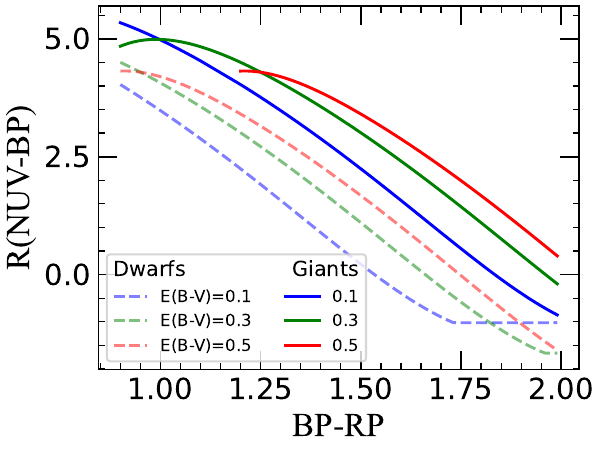}
\caption{R$_{\rm NUV-BP}$, as a function of BP$-$RP, for dwarfs (dashed lines) and giants (solid lines) at different extinction values.}
\label{fig:red-coef}
\end{figure}

In the python package of \citet{zhang2023}, the input temperatures could be replaced by $\rm BP-RP$ colors when unavailable, as adopted in this work.
During this process, we observe differences in the reddening coefficients between dwarfs and giants. We randomly select 10,000 sources in common with $\rm E(B-V)<0.5$ from LAMOST DR8 (\citealt{Luo2015}), Gaia EDR3, and GALEX GR6+7 to present the differences in the $\rm BP-RP$ vs. $\rm NUV-BP$ intrinsic color loci between giants and dwarfs in Figure \ref{fig:loci-diff},  
which illustrates that BP$-$RP colors differ between dwarfs and giants at the same NUV$-$BP color. This results in a modest bias in the reddening coefficients obtained from \citet{zhang2023} for giants, mainly due to the lack of giants in their UV sample. 
This means that the reddening coefficient will not be perfectly suitable for the giants, mainly because the giants are redder than the dwarfs in BP$-$RP at the same NUV$-$BP color. Therefore, we make an attempt to correct this bias by subtracting an empirical value of 0.3 mag from the BP$-$RP colors for giants when using the python package.  The result is shown in Figure \ref{fig:red-coef}.

All colors referred to hereafter are the intrinsic (de-reddened) colors unless otherwise noted.

\vskip 1cm
\subsection{$\rm [Fe/H]$- and $\rm M_G$-dependent Stellar Loci} \label{subsec:loci}
  
Similar to \citet{yuan2015metal}, we conduct a polynomial fitting for stellar loci, and employ a minimum $\chi^2$ technique to derive the metallicity for dwarfs and giants, respectively. Unlike \citet{yuan2015metal}, we consider the effects of both metallicities and absolute magnitude on stellar loci rather than metallicities alone (see APPENDIX \ref{sec:appendix}). A fourth-order three-dimensional polynomial with thirty-one free parameters is adopted to fit the relationship among NUV$-$BP, BP$-$RP color, [Fe/H], and $\rm M_G$ for both dwarfs and giants:
\begin{eqnarray}\label{eq1}
&(NUV-BP) = \nonumber \\
&p0\cdot X^4 + p1\cdot Y^4 + p2\cdot Z^4 + p3\cdot X^3\cdot Y \nonumber \\
&{}+ p4\cdot Y^3\cdot X + p5\cdot X^3\cdot Z + p6\cdot Z^3\cdot X \nonumber \\
&{}+ p7\cdot Z^3\cdot Y + p8\cdot Y^3\cdot Z + p9\cdot X^2\cdot Y^2 \nonumber \\
&{}+ p10\cdot X^2\cdot Z^2 + p11\cdot Z^2\cdot Y^2 + p12\cdot X^3 \nonumber \\
&{}+ p13\cdot Y^3 + p14\cdot Z^3 + p15\cdot X^2\cdot Y + p16\cdot Y^2\cdot X \nonumber \\
&{}+ p17\cdot X^2\cdot Z + p18\cdot Z^2\cdot X + p19\cdot Z^2\cdot Y \nonumber \\
&{}+ p20\cdot Y^2\cdot Z + p21\cdot X^2 + p22\cdot Y^2 + p23\cdot Z^2 \nonumber \\
&{}+ p24\cdot X\cdot Y + p25\cdot Y\cdot Z + p26\cdot X\cdot Z \nonumber \\
&{}+ p27\cdot X + p28\cdot Y + p29\cdot Z + p30,
\end{eqnarray}
where $X, Y, Z$ represents the BP$-$RP color, [Fe/H], and $\rm M_G$, respectively. The $\rm M_G$ here are calculated by:
\[
M_{G}= G - 10 - 5 \times  \log_{10}{\frac{1}{parallax}} -R_{G}\times E(B-V),
\]where the parallax values are the Gaia parallaxes corrected by \citet{Lindegren2021}, $\rm R_G$ values are from the python package of \citet{zhang2023}, and $\rm E(B-V)$ values are from the SFD98 redenning map.
During the fitting process, a 3$\sigma$-clipping procedure is performed to reject outliers. The fitting coefficients $p_i$ are provided in Table\ref{table1}. 
   
\setlength{\tabcolsep}{5mm}{
\begin{table}[htbp]
\footnotesize
\centering
\caption{Fitting Coefficients for the Dwarfs and Giants}
\begin{tabular}{c|l|l} 
\hline  
\hline
\textbf{Coeff.} & \textbf{Giants} & \textbf{Dwarfs} \\
\hline
p0    &   $-$0.7607     &    $+$5.8352     \\  
p1    &   $-$0.0207     &    $-$0.0231     \\  
p2    &   $-$0.0012     &    $-$0.0210     \\  
p3    &   $+$0.1987     &    $+$2.6037     \\  
p4    &   $-$0.0413     &    $+$0.0153    \\  
p5    &   $-$1.8229     &    $-$0.8798     \\  
p6    &   $-$0.0921     &    $+$0.4813     \\  
p7    &   $-$0.0068     &    $-$0.0089     \\  
p8    &   $-$0.0032     &    $-$0.0085     \\   
p9    &   $-$0.2876     &    $-$2.0478     \\  
p10   &   $-$0.9692     &    $-$1.3104     \\  
p11   &   $-$0.0000     &    $+$0.0198     \\  
p12   &   $-$5.9457     &    $-$11.2965    \\  
p13   &   $-$0.1228     &    $-$0.1001     \\  
p14   &   $+$0.1019     &    $+$0.0225     \\  
p15   &   $-$2.3209     &    $-$12.0546    \\  
p16   &   $+$0.3297     &    $+$2.8586     \\ 
p17   &   $+$5.8758     &    $+$9.1233     \\
p18   &   $+$2.1666     &    $-$3.7948     \\
p19   &   $+$0.0044     &    $+$0.1869     \\    
p20   &   $+$0.0142     &    $-$0.1947     \\
p21   &   $+$24.6050    &    $-$2.3854     \\
p22   &   $-$0.3129     &    $-$0.5011     \\
p23   &   $-$1.2202     &    $+$1.0914     \\  
p24   &   $+$3.8330     &    $+$13.9786    \\   
p25   &   $+$0.1420     &    $-$1.0543     \\
p26   &   $-$6.2672     &    $+$10.1043    \\  
p27   &   $-$20.7169    &    $-$8.6359     \\
p28   &   $-$0.9627     &    $-$1.7705     \\
p29   &   $+$2.2438     &    $-$5.4122     \\  
p30   &   $+$9.5277     &    $+$8.6614     \\
\hline
\end{tabular}
\label{table1}
\end{table}}

\begin{figure*}[htbp]
\centering
\includegraphics[width=14cm]{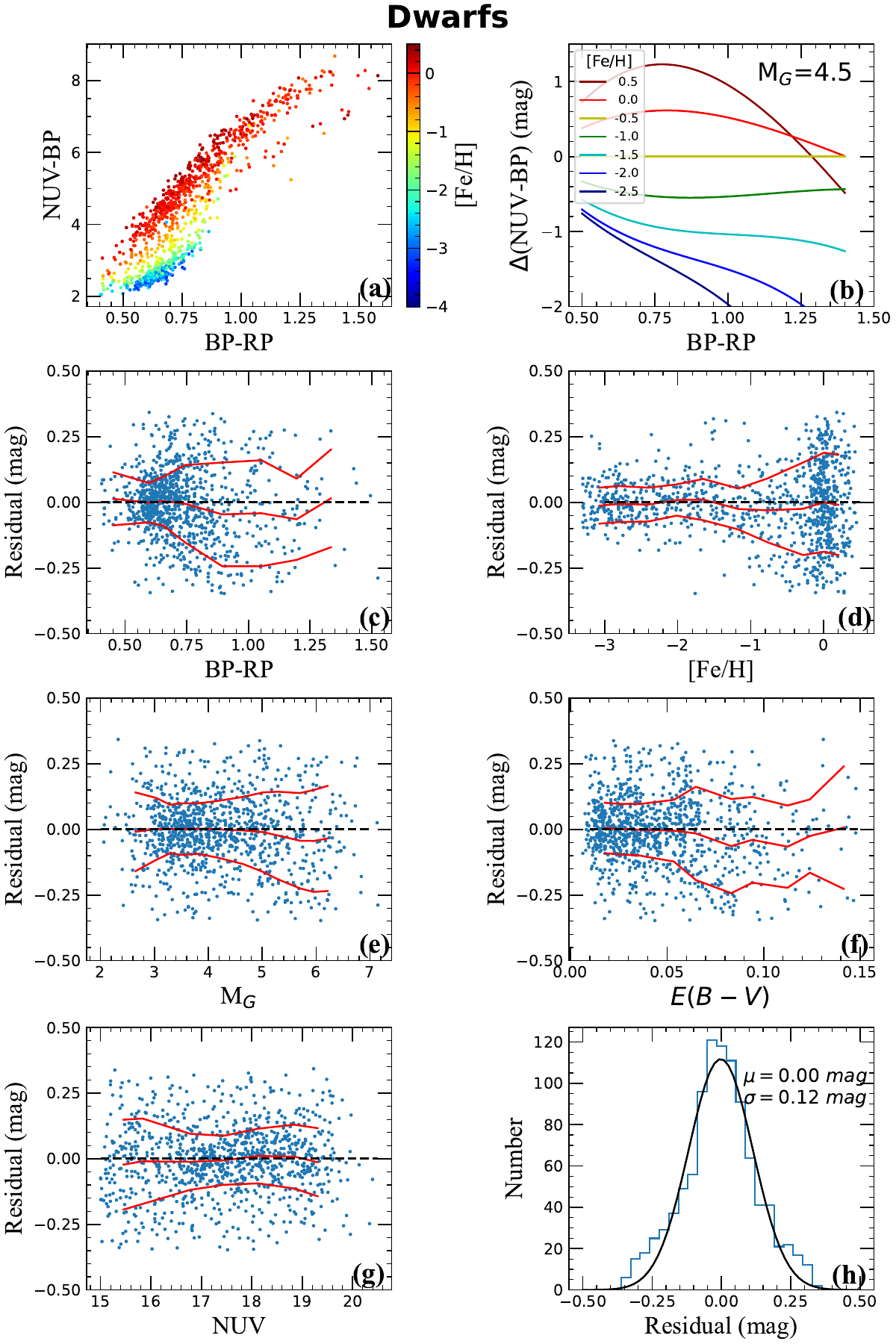}
\caption{Panel (a): Distributions of the dwarf training sample in the NUV$-$BP vs. BP$-$RP plane (stellar loci), color-coded by [Fe/H] as shown in the right color bar. Panel (b): When $\rm M_G = 4.5$, the variations of stellar loci for different metallicities (as shown in the label) relative to the one at [Fe/H] = $-$0.5. Panels (c--g): Fitting residuals as a function of BP$-$RP color, [Fe/H], $\rm M_G$, E(B$-$V), and NUV, respectively. The red-solid lines indicate the median values and standard deviations. The black-dashed line indicates the zero level. Panel (h): Histogram distribution of the fitting residuals, with the Gaussian fitting profile over-plotted in black.
}
\label{fig:fit-dwarf}
\end{figure*}

\begin{figure*}[htbp]
\centering
\includegraphics[width=14cm]{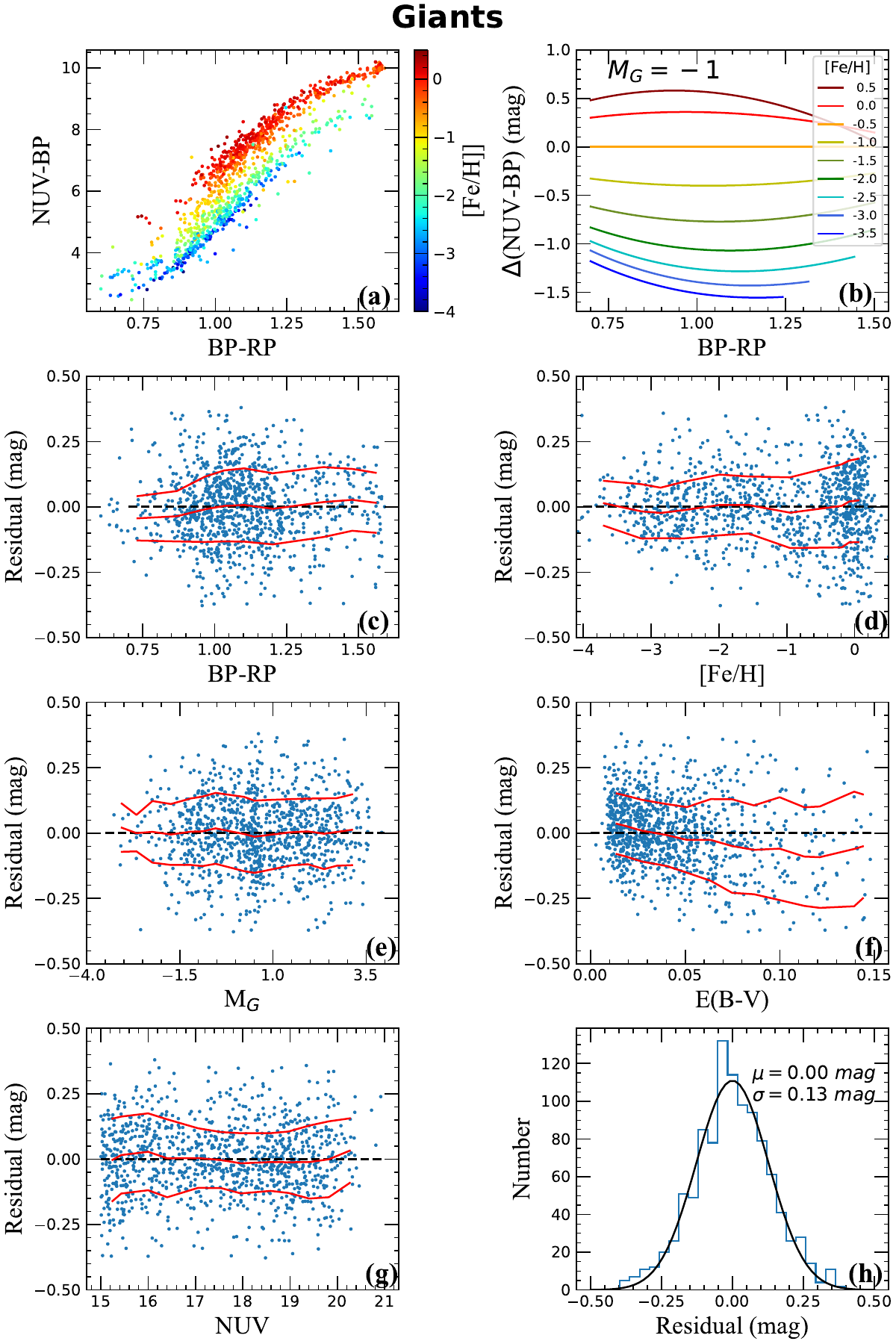}
\caption{Same as Figure \ref{fig:fit-dwarf}, but for the giant training sample.}
\label{fig:fit-giant}
\end{figure*}

The fitting results for the dwarf training sample are shown in Figure \ref{fig:fit-dwarf}. The distribution of the dwarf training sample in the NUV$-$BP vs. BP$-$RP plane is plotted in panel (a), color-coded by [Fe/H]. From inspection, there are clear dividing lines between sources with different metallicities.
Panel (b) shows the variations of the stellar locus at $\rm M_G=4.5$ for different metallicities relative to the one at [Fe/H] = $-$0.5. A 1\,dex change in metallicity corresponds to an approximately 1 mag change in (NUV$-$BP) color for Solar-type stars, which is about 2--3 times larger than the change in the $\rm v-BP$ color (\citealt{Huang2023}), about 5 times larger than the change in the $u-g$ color (\citealt{yuan2015FGK}), and about 200 times larger than the change in the BP$-$G color (\citealt{Xu2022gaia-metal}). 
This can be attributed to the prevalence of strong metallic absorption lines in the NUV-band, leading to heightened sensitivity to metallicity, as expected. 
The fitting residuals, as a function of BP$-$RP, [Fe/H], $\rm M_G$, E(B$-$V), and NUV are plotted in panels (c), (d), (e), (f), and (g), respectively. No trends are found. Additionally, it is noted that the dispersion differs between metal-poor and metal-rich stars. This is due to the stellar loci being more sensitive to metallicity for metal-rich stars, resulting in greater dispersion for the NUV-BP fitting residuals.
Panel (h) plots the histogram distribution of the fitting residuals. A Gaussian profile is fitted to the distribution, resulting in a narrow peak centered at 0.00 mag with a $\sigma$ of 0.13 mag. 
The fitting results of the giant training sample are similar, and shown in Figure \ref{fig:fit-giant}.

Based on the [Fe/H]- and $\rm M_G$-dependent stellar loci,
we employ a minimum $\chi^2$ technique to estimate the photometric metallicities from 
NUV$-$BP,  BP$-$RP, and absolute magnitude $\rm M_G$. The $\chi^2$ is defined as:
\begin{eqnarray}\label{eq3}
&&\chi^2([\rm Fe/H]) = \nonumber \\ 
&&\frac{[\rm C^{int}_{NUV-BP} - C^{int}_{NUV-BP}(BP-RP, [Fe/H], M_G)]^2}{\sigma_{\rm NUV}^2 +\sigma_{\rm BP}^2},
\end{eqnarray}
where $\rm C_{NUV-BP}^{int}$ is the intrinsic NUV$-$BP color, $\rm C^{int}_{NUV-BP}(BP-RP, [Fe/H])$ represents the predicted intrinsic NUV$-$BP color based on the metallicity- and $\rm M_G$-dependent stellar loci, and $\sigma_{\rm NUV}^2$ and $\sigma_{\rm BP}^2$ represent the magnitude uncertainties in NUV and BP, respectively. 
Indeed, $\sigma_{\rm BP}^2$ could be safely ignored compared to $\sigma_{\rm NUV}^2$. We utilize a brute-force algorithm to determine the optimal [Fe/H] value for each source. For a given dwarf, the [Fe/H] value is varied from $-$4.0 to +1.0 with steps of 0.01\,dex, resulting in 501 $\chi^2$ values. For a given giant, the [Fe/H] value is varied from $-$5.0 to +1.0 with steps of 0.01\,dex. 
The minimum $\chi^2$ among these values corresponds to the optimal [Fe/H] value. As expected, almost all our minimum chi-square values are very close to zero. Secondary minima will only occur when crossings between loci of different [Fe/H] happen. It is clear that there are almost no crossings in most cases, as evidenced in panel (b) of Figures\,\ref{fig:result-dwarf} and \ref{fig:result-giant}. The only exception is the stellar loci for very metal-rich and red stars. In such cases, if a secondary minimum is present, we give preference to the smaller [Fe/H] value. Moreover, we don't provide uncertainties of our [Fe/H] estimates, considering the probable significant under-estimation of photometric errors in the NUV band. The 
real uncertainties can be better characterized by comparing with 
external catalogs.

\vskip 1cm
\section{Results and discussion} \label{sec:resu-disc}

We employed the above model and technique to determine the photometric-metallicity 
estimates, denoted as $[\rm Fe/H]_{GALEX}$, based on the NUV$-$BP,  BP$-$RP, and absolute magnitude $\rm M_G$ obtained from Gaia EDR3 and GALEX GR6+7 AIS datasets. In this section, we test and verify the accuracy and precision of our approach.
  
\subsection{Results for the Training Sample} \label{subsec:result}
\begin{figure*}[htbp]
\centering
\includegraphics[width=14cm]{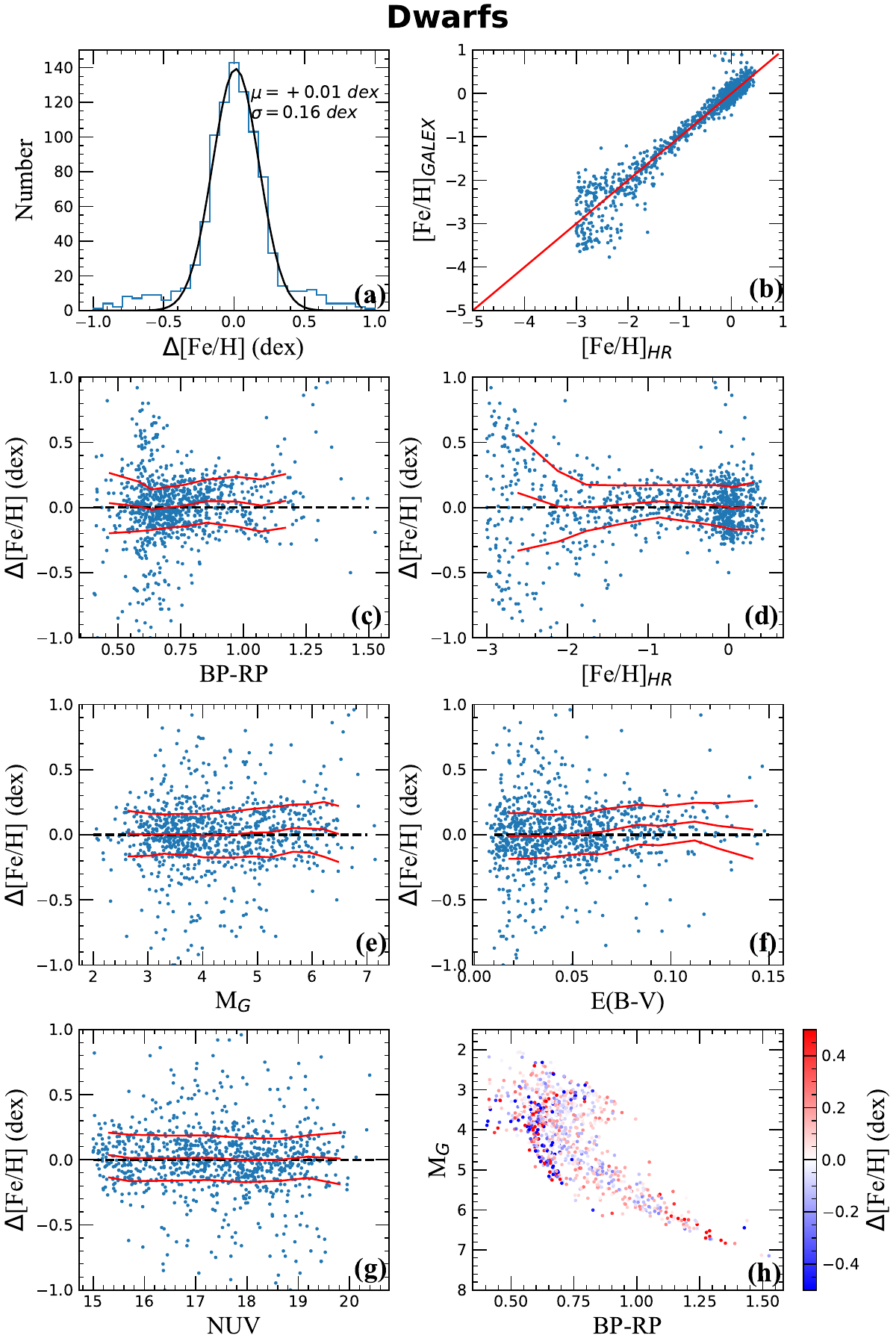}
\caption{Photometric metallicities for the dwarf training sample. Panel (a): Histogram distribution of the residuals $\Delta [\rm Fe/H]$, with the Gaussian fitting profile over-plotted in black. Panel (b): Comparison of photometric-metallicity estimates from GALEX and Gaia data with high-resolution spectroscopic metallicities from SAGA or PASTEL. Panels (c--g): $\Delta[\rm Fe/H]$ as a function of BP$-$RP color, $\rm [Fe/H]_{HR}$, $\rm M_G$, E(B$-$V), and NUV, respectively. The red-solid lines indicate the median values and standard deviations. The black-dashed line indicates the zero level. Panel (h): The color-magnitude diagram, color-coded by $\Delta [\rm Fe/H]$.}
\label{fig:result-dwarf}
\end{figure*}

\begin{figure*}[htbp]
\centering
\includegraphics[width=14cm]{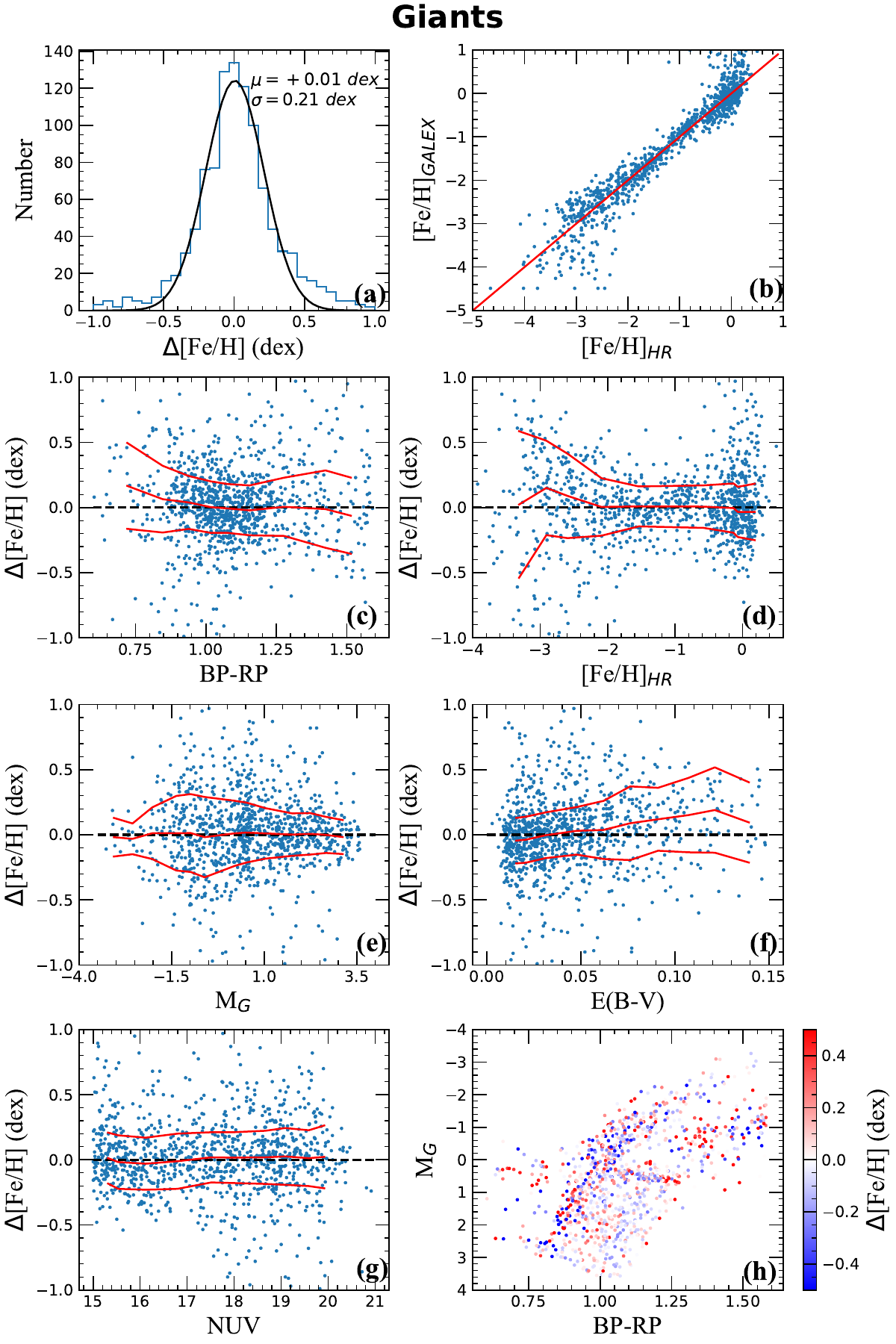}
\caption{Same as Figure \ref{fig:result-dwarf}, but for the giant training sample.}
\label{fig:result-giant}
\end{figure*}

The photometric metallicities for the dwarf and giant training samples are presented in Figures \ref{fig:result-dwarf} and \ref{fig:result-giant}, respectively.
The residuals $\Delta$[Fe/H], defined as $\Delta [\rm Fe/H] = [\rm Fe/H]_{GALEX} - [\rm Fe/H]_{\rm HR}$, is used to assess the precision of our results here, where $[\rm Fe/H]_{\rm HR}$ is [Fe/H] from high-resolution spectroscopy (PASTEL and SAGA).
Panel (a) shows a histogram distribution of the $\Delta [\rm Fe/H]$. The standard deviation of $\Delta$[Fe/H] for the dwarf training sample is 0.16\,dex, with a mean value of +0.01\,dex. For the giant training sample, the standard deviation is 0.21\,dex, with a mean value of +0.01\,dex. Panel (b) shows the correlation between $[\rm Fe/H]_{\rm GALEX}$ and $[\rm Fe/H]_{\rm HR}$. Strong one-to-one correlations are found both for [Fe/H] down to [Fe/H] $\sim-3.0$ for the dwarfs and to [Fe/H] $\sim-4.0$ for the giants.  Panels (c), (d), (e), (f), (g) show the distributions of $\Delta$[Fe/H] as a function of BP$-$RP color, $[\rm Fe/H]_{\rm HR}$, $\rm M_G$, E(B$-$V), and NUV, respectively. No discernible trends are found. 
    
In panel (b) of Figure \ref{fig:result-dwarf}, at $\rm [Fe/H]<-2.0$, the dispersion for dwarfs between $[\rm Fe/H]_{\rm GALEX}$ and $[\rm Fe/H]_{HR}$ increases, corresponding to the sources in panel (h), whose color slightly deviates from $\Delta[\rm Fe/H]= 0.0$\,dex. This is due to the fact that lower metallicity corresponds to a lower sensitivity of the stellar locus. Consequently, estimating precise photometric metallicities for very metal-poor (VMP; [Fe/H] $\leq -2.0$) and extremely metal-poor (EMP; [Fe/H] $\leq -3.0$) stars becomes challenging. This challenge is also evident for giants, though with slightly improved performance, as seen in panel (b) of Figure \ref{fig:result-giant}.
    
Panel (h) of Figure \ref{fig:result-dwarf} and Figure \ref{fig:result-giant} plots $\Delta$[Fe/H] in the color-magnitude diagram.
For dwarfs, most $\Delta[\rm Fe/H]$ values are very close to zero. A small region with slight deviations from $\Delta[\rm Fe/H]=0$\,dex is derived from the larger dispersion in the most metal-poor stars. Overall, $\Delta [\rm Fe/H]$ has no significant structure associated with $\rm M_G$, which is attributed to our consideration of $\rm M_G$ of our model.
There is also no significant structure associated with $\rm M_G$ for giants. For some individual outliers, it is not meaningful to consider them due to the scarcity of these sources. We consider the outliers in our comparison with the test sample below.

\begin{figure*}[htbp]
\centering
\includegraphics[width=14cm]{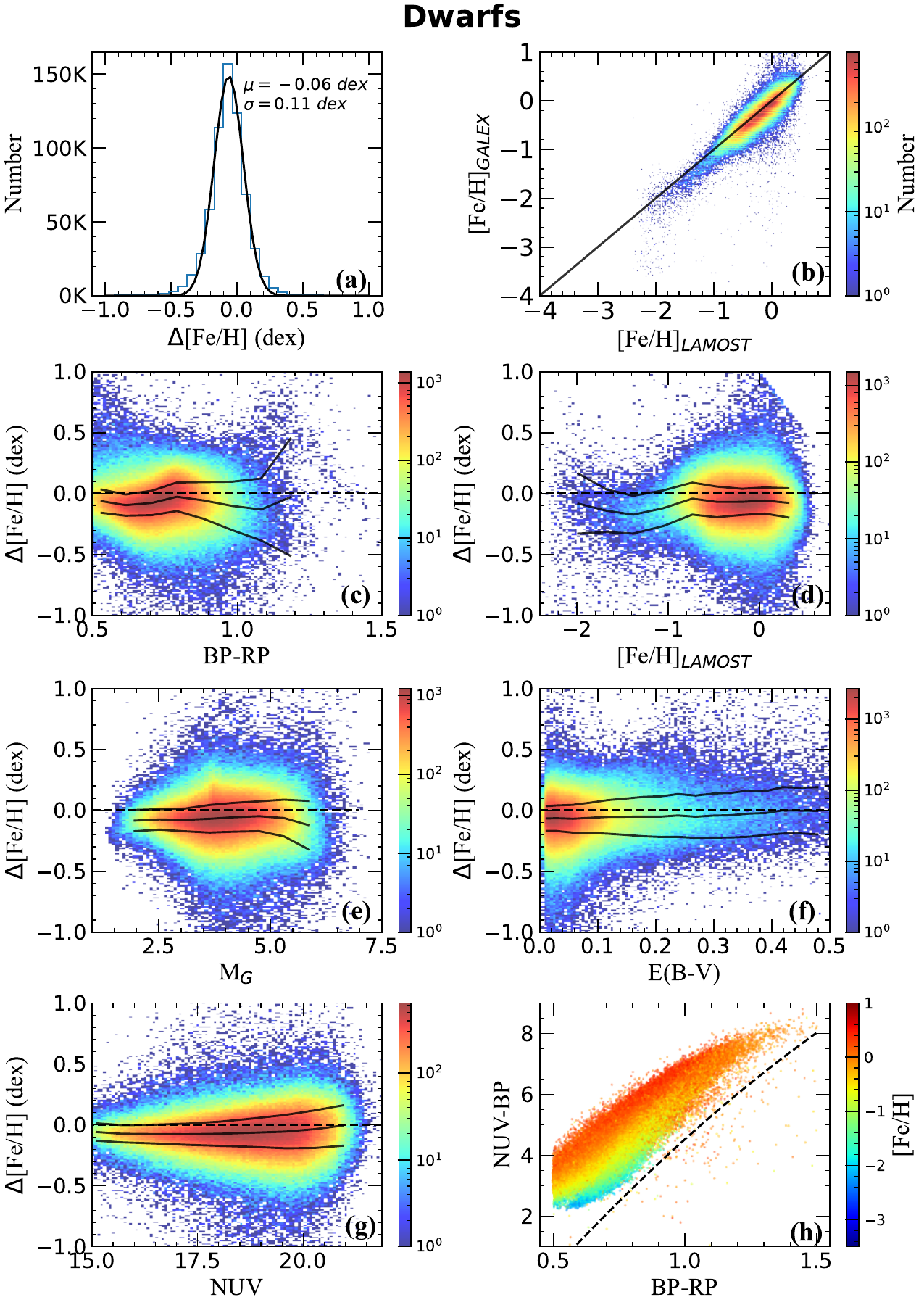}
\caption{Test results for 633,807 dwarfs. Panel (a): Histogram distribution of the residuals $\Delta [\rm Fe/H]$, with the mean and standard deviation values labeled. Panel (b): $[\rm Fe/H]_{GALEX}$ as a function of $[\rm Fe/H]_{LAMOST}$, color-coded by number density. Panels (c), (d), (e), (f), (g): $\Delta [\rm Fe/H]$ as a function of BP$-$RP color, $[\rm Fe/H]_{LAMOST}$, $\rm M_G$, E(B$-$V), and NUV, respectively. The black-solid lines indicate the median values and standard deviations. The black-dashed line indicates the zero level.  Panel (h): Distributions of the dwarfs in the (NUV $-$ BP) vs. (BP $-$ RP) plane, color-coded by [Fe/H] values from LAMOST, as indicated by the right color bar. The expression for the black-dashed line is: $(\rm NUV - BP) < -1.56 \times (BP - RP)^2 + 10.87 \times (BP - RP) - 4.78$ for dwarfs. The $[\rm Fe/H]_{GALEX}$ values of stars below the black-dashed line are considered unreliable and assigned `flag=1'.}

\label{fig:lamo-dwarf}
\end{figure*}   

\begin{figure*}[htbp]
\centering
\includegraphics[width=14cm]{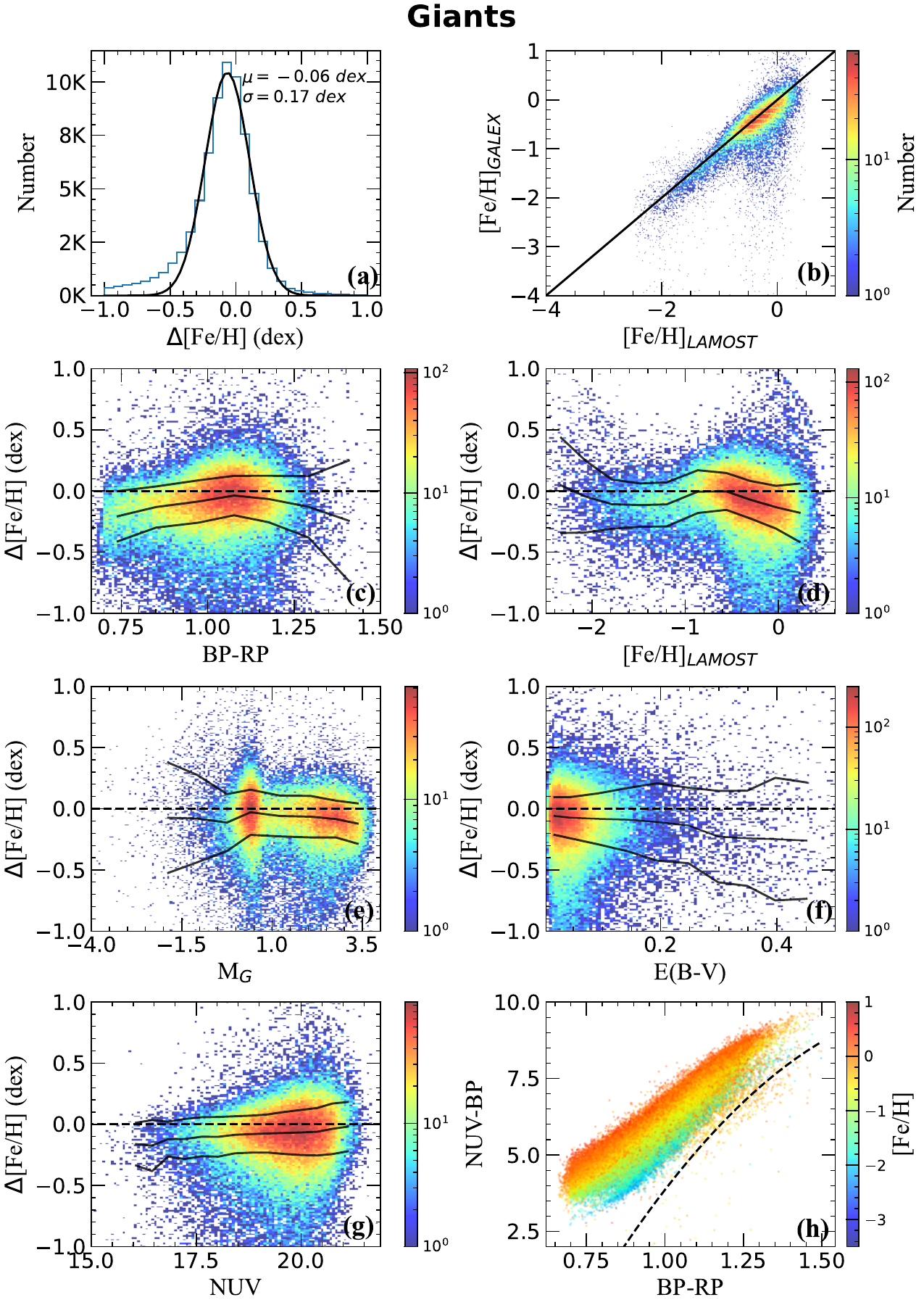}
\caption{Same as Figure \ref{fig:lamo-dwarf}, but for 72,911 giants. The expression for the black-dashed line is: $\rm (NUV-BP) < -7.20 \times (BP-RP)^2 + 27.69 \times (BP-RP) -16.63$. }
\label{fig:lamo-giant}
\end{figure*}

\begin{figure*}[htbp]
\centering
\includegraphics[width=14cm]{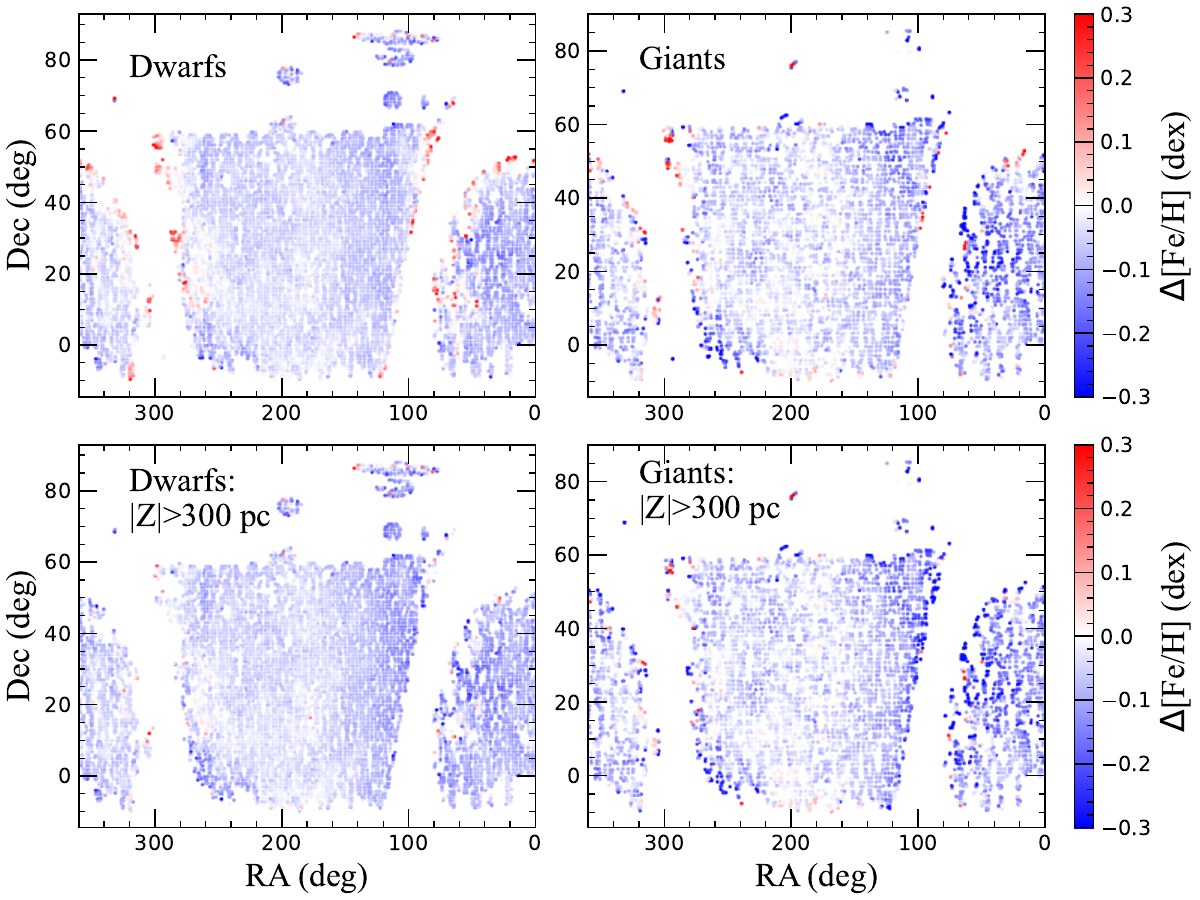}
\caption{The spatial distribution (RA and Dec) of the residuals $\Delta$[Fe/H]. The top row of panels are for the test-with-LAMOST sample of 633,807 dwarfs and 72,911 giants, respectively. The bottom row of panels are the sub-sample of these stars with $|Z|>300$ pc (441,141 dwarfs and 64,696 giants, respectively.)}
\label{fig:ra-dec}
\end{figure*}

\vskip 1cm
\subsection{External Tests with LAMOST DR8} \label{subsec:test-dr8}
    
In order to verify the accuracy of our model, some external checks are performed in this section by comparing with spectroscopic-metallicity measurements from LAMOST DR8. First, we cross-match the LAMOST DR8 data with GALEX GR6+7 AIS and Gaia EDR3. After application of the following cuts:
\begin{enumerate}
\item Signal-to-noise ratio for the g-band of the LAMOST spectra $\rm SNR_g >20$
\item Galactic latitude $|b|>10^\circ$ and $\rm E(B-V)\leq0.5$
\item Sources with $\rm 10<G\leq16$ and within the limits of BP$-$RP color and $\rm phot\_bp\_rp\_excess\_factor$ (the same as \citealt{Xu2022gaia-metal}),
\end{enumerate}
we retain a sample with 633,807 dwarfs and 72,911 giants. The comparison results are displayed in Figures\, \ref{fig:lamo-dwarf}, \ref{fig:lamo-giant}, and \ref{fig:ra-dec}. Note that there are small metallicity offsets between the LAMOST and training sample: $-0.07$\,dex for dwarfs, and $-0.06$\,dex for giants. The residual $\Delta [\rm Fe/H]$ is defined as: $\Delta [\rm Fe/H]=[\rm Fe/H]_{GALEX}- [\rm Fe/H]_{LAMOST}$. 
    
For the dwarfs in Figure \ref{fig:lamo-dwarf}, we fit the histogram distribution of the residuals $\Delta[\rm Fe/H]$ with a Gaussian profile, and obtain a small standard deviation of only 0.11\,dex, centered at $-$0.06\,dex, as shown in panel (a). Panel (b) shows the relationship between $[\rm Fe/H]_{GALEX}$ and $[\rm Fe/H]_{LAMOST}$. Most sources exhibit good agreement between $\rm [Fe/H]_{GALEX}$ and $\rm [Fe/H]_{LAMOST}$, except for a tiny fraction of sources located at the bottom right, which could be due to stellar activity in the NUV-band. Note that the outliers near $\rm [Fe/H]_{LAMOST}=-2.0$ is due to the minimum metallicity of the LAMOST estimates, $\sim-2.5$. From inspection of panels (c--g), we conclude that the $\Delta [\rm Fe/H]$ values exhibit no discernible trends with BP$-$RP color, $[\rm Fe/H]_{LAMOST}$, $\rm M_G$, E(B$-$V), and NUV. 
Panel (h) shows the distributions of the 633,807 dwarfs in the NUV$-$BP vs. BP$-$RP plane, color-coded by $\rm [Fe/H]_{LAMOST}$. The black-dashed line is used to separate the outliers. From inspection, there are some cooler metal-rich stars located below the black-dashed line, we assume due to stellar activity. These sources would be miss-classified as metal-poor stars in our model. Therefore, we add a `flag' parameter to mark these outliers. The `flag=1' represents outliers likely caused by stellar activity; stars with`flag=0' are not.
    
For the giants in Figure \ref{fig:lamo-giant}, the dispersion of the residuals 
$\Delta[\rm Fe/H]$ is 0.17\,dex, with a mean value of $-$0.06\,dex, as shown in panel (a). As indicated by the relationship between $[\rm Fe/H]_{GALEX}$ and $[\rm Fe/H]_{LAMOST}$ in panel (b), the majority of sources also exhibit good agreement between $\rm [Fe/H]_{GALEX}$ and $\rm [Fe/H]_{LAMOST}$. However, there are more outliers located at the bottom right compared to the dwarfs. This could be due to more stellar activity in cool giants, leading to a decrease in the NUV magnitude and lower $[\rm Fe/H]_{GALEX}$ values. In panels (c--g), there are some subtle trends in the residuals worth discussing. In panel (c), the median value of $\Delta [\rm Fe/H]$ is slightly smaller at bluer and redder BP$-$RP colors, which is mainly attributed to the scarcity of bluer and redder $\rm BP-RP$ sources in our training sample. The result of panel (d) is similar to that shown in panel (b). As for the slope observed in metal-rich stars, it mainly results from the  systematic differences between the metallicities from our training sample and those from LAMOST. This is evidenced by the similar slope found for the giant stars in common between the training sample and LAMOST DR8.
Panel (e) shows that the errors are larger for intrinsically brighter stars. Panel (f) shows that both the median differences and errors increase as $\rm E(B-V)$ increases. While the errors display very weak dependence on NUV magnitude, as shown in panel (g), and also seen in the dwarfs.
Panel (h) shows the distributions of the 72,911 giants in the NUV$-$BP vs. BP$-$RP plane, color-coded by $\rm [Fe/H]_{LAMOST}$. The black-dashed line is used to separate the outliers.

Figure \ref{fig:ra-dec} shows the spatial distribution of the residuals $\Delta$[Fe/H]. Due to the over-estimation of extinction values in the SFD98 reddening map close to the Galactic plane, there is an over-estimation of [Fe/H] in our model for the stars in this region, as shown in the top row of
panels (especially in dwarfs). 
For a typical star of $\rm E(B-V)$ = 0.3, [Fe/H] = 0, 
and BP$-$RP = 1, [Fe/H] will be over-estimated by 
approximately 0.1\,dex for dwarfs and 0.2\,dex for giants if $\rm E(B-V)$ is over-estimated by 0.05 mag.
We can eliminate this over-estimation from SFD98 by applying a $|Z|>300$\,pc cut, as shown in the bottom row of panels.

Based on these test results, most metallicities estimated by our model exhibit good agreement with those from LAMOST DR8. The outliers are primarily likely active stars, especially for giants. We have added a `flag' parameter to label these active stars (with `flag=1'). Additionally, we can eliminate sources with over-estimation of reddening from SFD98 by applying a $|Z| > 300$\,pc cut.

\subsection{Tests with Other Catalogs} \label{subsec:addi-test}
\begin{figure*}[htbp]
\centering
\includegraphics[width=20cm]{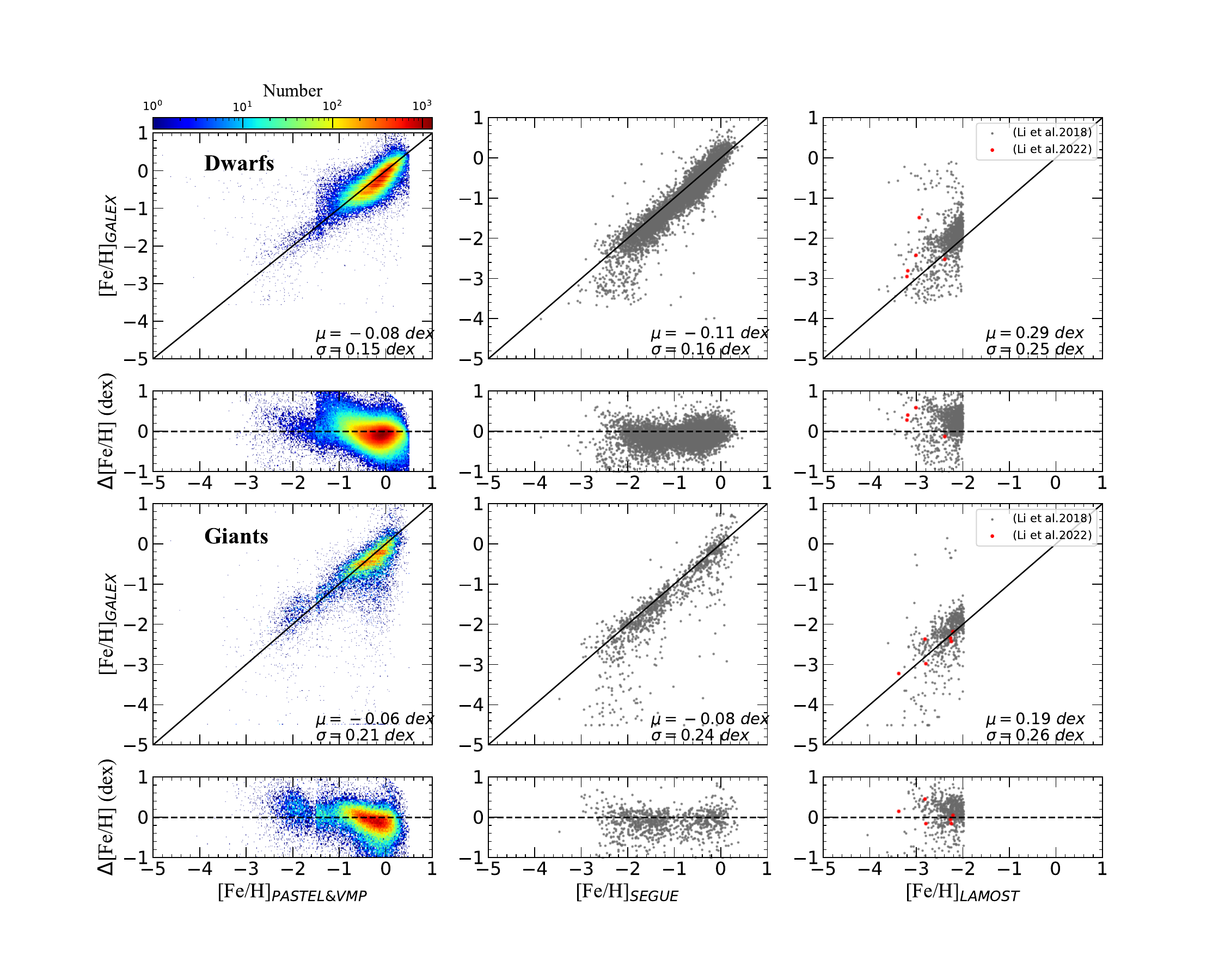}
\caption{From left to right, the columns of panels represent comparisons with the LAMOST DR8 value-added catalog of \citet{Wang_2022}, the SEGUE DR12 (\citealt{sdssdr12}) catalog, and the VMP catalogs of \citet{Li_2018} and \citet{Li_2022}, respectively. The left column of panels is color-coded by the source number density. In the right column of panels, the gray and red dots are from \citet{Li_2018} and \citet{Li_2022}, respectively. The dwarfs are plotted in the top 
row of panels and the giants in the bottom row of panels. The mean and standard deviation values of the residuals $\Delta[\rm Fe/H]$ are labeled in the lower-right of each panel.}
\label{fig:test-add}
\end{figure*}

Comparisons with other independent stellar-metallicity measurements (spanning a wide range of [Fe/H] values) are presented in this section. Four catalogs are used here: the value-added catalog of LAMOST DR8 (\citealt{Wang_2022}), the SEGUE DR12 (\citealt{sdssdr12}) catalog, the catalog of $\sim10,000$ VMP stars from LAMOST DR3 (\citealt{Li_2018}), and the 400 VMP metal-poor stars studied with LAMOST and Subaru (\citealt{Li_2022}).
To perform the comparisons, we cross-match these catalogs with the Gaia EDR3 and GALEX GR6+7 AIS catalogs, respectively. We retain the sources after applying the criteria similar to those used for the LAMOST DR8 test sample. The results are presented in Figure \ref{fig:test-add}, with the top two panels for dwarfs and the bottom two panels for giants. In each of the upper sub-panels of Figure \ref{fig:test-add}, we can see the comparison of metallicities estimated by our model with those from other catalogs, with a $\Delta$[Fe/H] as a function of metallicities from other catalogs shown in the lower sub-panels.
    
In the left column of panels, we show the comparison with the LAMOST DR8 value-added catalog. We utilized the $[\rm Fe/H]_{PASTEL}$ and $[\rm Fe/H]_{VMP}$ values from the LAMOST DR8 value-added catalog. The $[\rm Fe/H]_{PASTEL}$ represents the value of [Fe/H] with the neural network using LAMOST–PASTEL sample as a training set. The $[\rm Fe/H]_{VMP}$ represents the improved [Fe/H] values of VMP candidates. More details can be found in \citealt{Wang_2022}. $[\rm Fe/H]_{PASTEL\& VMP}$ represents $[\rm Fe/H]_{PASTEL}$ (for $[\rm Fe/H]>-1.5$) and $[\rm Fe/H]_{VMP}$ (for $[\rm Fe/H]<-1.5$) from \citet{Wang_2022}. The residuals between the $[\rm Fe/H]_{GALEX}$ and $[\rm Fe/H]_{PASTEL\& VMP}$ were fitted with a Gaussian profile. For dwarfs, we obtain a dispersion of 0.15\,dex centered at $-$0.08\,dex. For giants, a dispersion of 0.21\,dex centered at $-$0.06\,dex is obtained.
The middle column of panels displays the comparison with SEGUE DR12. For dwarfs, we obtain a dispersion of 0.16\,dex centered at $-$0.11\,dex. 
For giants, a dispersion of 0.25\,dex centered at $-$0.07\,dex is obtained. 
The right column of panels displays the comparison with the catalogs of \citet{Li_2018,Li_2022}. For dwarfs, we obtain a dispersion of 0.25\,dex centered at +0.29\,dex. For giants, the dispersion is 0.26\,dex centered at +0.19\,dex. From inspection, except for a small fraction of outliers, our [Fe/H] measurements demonstrate good consistency with other catalogs for most sources, encompassing both metal-rich and metal-poor stars. As for the left column, there is a similar systematic difference compared to the test results with LAMOST DR8.
    
There are a few sources with $\rm [Fe/H]_{GALEX}$ values larger than $\rm [Fe/H]_{obs}$ located in the upper part of the right panels of Figure \ref{fig:test-add}. Their locations in the color-color diagram are close to the metal-rich stellar loci, probably due to their large photometric errors. 
In addition, there is an under-estimation of [Fe/H] in our model for a portion of sources with $[\rm Fe/H]_{GALEX}$ values lower than $-$3.0, as can be mainly seen in the bottom middle panel of Figure \ref{fig:test-add}. This is because the metallicity dependence decreases monotonically as NUV$-$BP color decreases at the same BP$-$RP color. Thus, between the over-estimates or under-estimates of NUV$-$BP colors caused by photometric errors, the under-estimates lead to a larger under-estimation in the calculation of photometric metallicities. Overall, from these tests, [Fe/H] estimates from our model are reliable even for metallicity down to [Fe/H] $\sim-$3.5.

\subsection{Tests With Star Clusters} \label{subsec:test-clusters}

\begin{figure*}[htbp]
\centering
\subfigure{
\includegraphics[width=7cm]{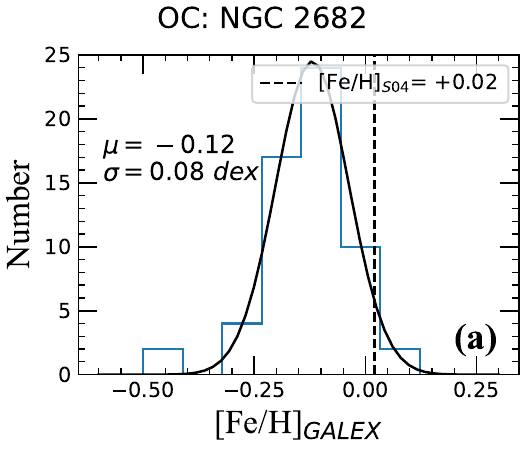}
}
\subfigure{
\includegraphics[width=7cm]{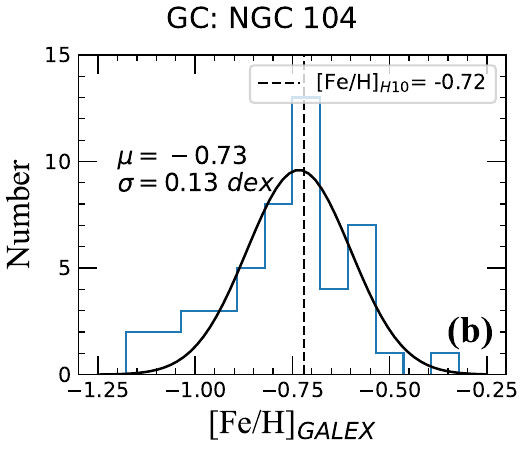}
}
\subfigure{
\includegraphics[width=7cm]{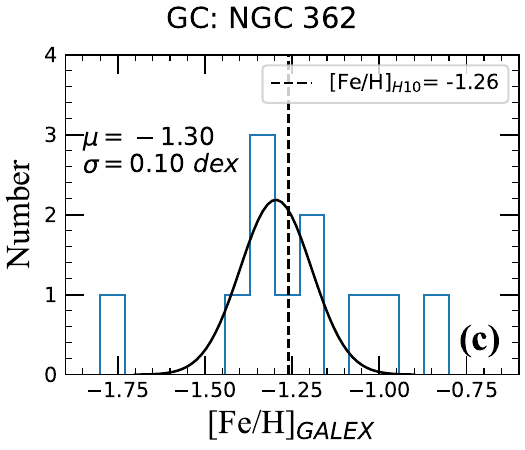}
}
\subfigure{
\includegraphics[width=7cm]{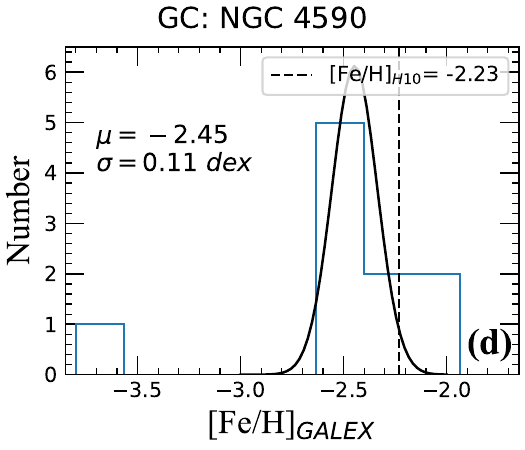}
}

\caption{Histograms of [Fe/H]$_{\rm GALEX}$ for the open cluster NGC\,2682 (panel (a)) and the globular cluster NGC\,104 (panel (b)), NGC\,362 (panel (c)), NGC\,4590 (panel (d)), with the Gaussian fitting profile over-plotted in black, and the mean and standard deviation values labeled. The values [Fe/H]$\rm _{S04}$ and [Fe/H]$\rm _{H10}$ are referenced from \citet{Salaris_2004} and \citet{harris_2010}, respectively, and are indicated with dashed vertical lines.  See text for values of the biweight estimates of location and dispersion.
}
\label{fig:test-clusters}
\end{figure*}

This section shows test results with star clusters. 
We cross-match the NGC\,2682 open cluster catalog of \cite{Cantat-Gaudin_2018}  with the Gaia EDR3 and GALEX GR6+7 AIS catalogs, and  retain the sources after applying the criteria similar to those used for the LAMOST DR8 test sample. The result is presented in panel (a) of Figure \ref{fig:test-clusters}.  A mean value of [Fe/H]$\rm _{GALEX} = -0.12$, and a dispersion of 0.08\,dex, are obtained for our photometric metallicity [Fe/H]$\rm _{GALEX}$. 
Note that the mean value of [Fe/H]$\rm _{GALEX}$ is slightly smaller than the literature value of \cite{Salaris_2004}, [Fe/H]$\rm_{S04} = +0.02$. The biweight estimates (see, e.g., \citealt{Beers1990}) of central location and scale (dispersion) are $-$0.12 and 0.09\,dex, respectively.

Similarly, following the work of \cite{Huang2019}, we select member stars of the globular clusters NGC\,104, NGC\,362, and NGC\,4590, and cross-match with the  Gaia EDR3 and GALEX GR6+7 AIS catalogs.\footnote{The relatively low numbers of stars selected for globular clusters is a result of 1) globular clusters are usually distant and compact and 2) the large errors associated with the GALEX NUV measurements for dense fields.}  The result is presented in the panels (b),(c), and (d) of Figure \ref{fig:test-clusters}. For NGC\,104, a dispersion of 0.13\,dex, centered at [Fe/H]$\rm _{GALEX} = -0.73$, is obtained for our photometric metallicity estimate. The mean value  closely matches [Fe/H]$\rm _{H10} = -0.72$, the latter referring to the metallicity value from \cite{harris_2010}. The biweight estimate of the location and dispersion are $-$0.76 and 0.22\,dex, respectively. For NGC\,362, the mean value $\rm [Fe/H]_{GALEX}=-1.30$, with a dispersion of 0.10\,dex, also closely matches the literature value of [Fe/H]$\rm _{H10} = -1.26$. The biweight estimate of the location and dispersion are $-$1.27 and 0.24\,dex, respectively. For NGC\,4590, the mean value of $\rm [Fe/H]_{GALEX}=-2.45$, with a dispersion of 0.11\,dex, reasonably matches the literature value of [Fe/H]$\rm _{H10} = -2.23$. The biweight estimate of the location and dispersion are $-$2.36 and 0.22\,dex, respectively.  Note that  the biweight estimates may be superior for the globular clusters, given their resistance to outliers and the low number of member stars involved.

\subsection{The Influence of Carbon Enhancement} \label{subsec:test-C-enhancement}

Photometric-metallicity measurements are often over-estimated for VMP/EMP stars, due to contamination of the blue narrow/medium-band filters by molecular carbon bands such as CN (e.g., \citealt{Hong2023}, \citealt{Huang2023}). In this subsection, we check whether the strong molecular carbon bands strongly influence the photometric-metallicity estimates found by use of the NUV-band. We cross-match the medium- and high-resolution spectroscopic catalog of VMP/EMP stars with available [Fe/H] and [C/Fe] measurements compiled by \citet{Hong2023} with our Gaia EDR3 and GALEX GR6+7 AIS catalog. After applying metallicity cuts of $\rm -4<[Fe/H]_{spec} \leq -2, [Fe/H]_{GALEX}>-4$, and $\rm flag=0 $, we obtain a sample of 231 stars in common.  

\begin{figure*}[htbp]
\centering
\includegraphics[width=18cm]{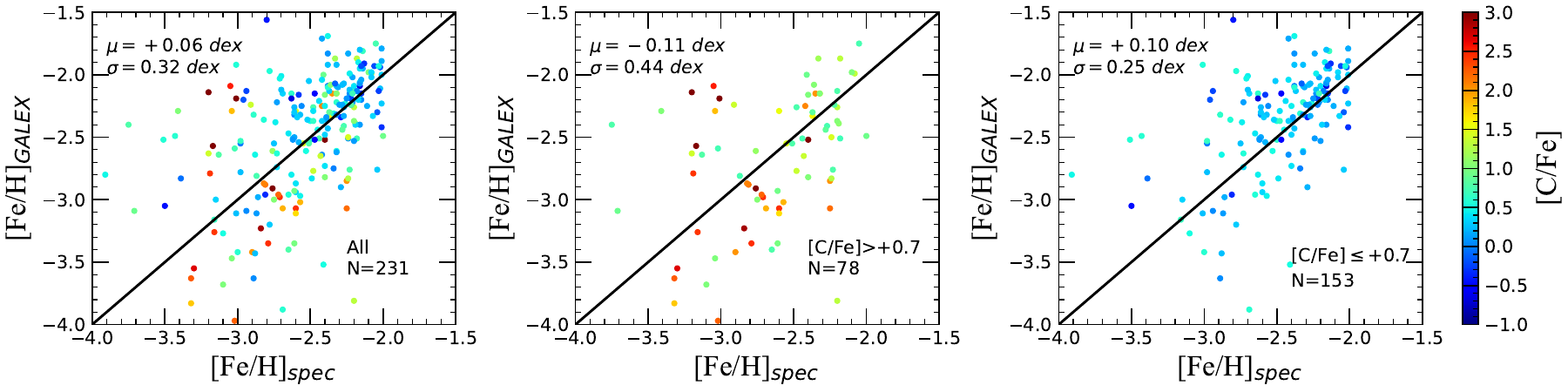}
\caption{Comparisons of $[\rm Fe/H]_{GALEX}$ with spectroscopic measurements for VMP/EMP stars (taken from the compilation of \citealt{Hong2023}), color-coded by [C/Fe]. From left to right, the panels show the full sample, the sub-sample of stars with [C/Fe] $>$ +0.7, and the sub-sample of stars with [C/Fe] $\leq$ +0.7, respectively. The mean and standard deviation values of the residuals $\Delta[\rm Fe/H]$ are labeled in the upper-left of each panel. Note that, in all three panels, there are stars with $[\rm Fe/H]_{GALEX}$ that straddle the black-solid one-to-one lines; they are \textit{not} uniformly biased high due to the effect of carbon, as has been seen to occur for several previous photometric-metallicity techniques (see text).}
\label{fig:test-cfe}
\end{figure*}

\begin{figure}[htbp]
\centering
\includegraphics[width=8cm]{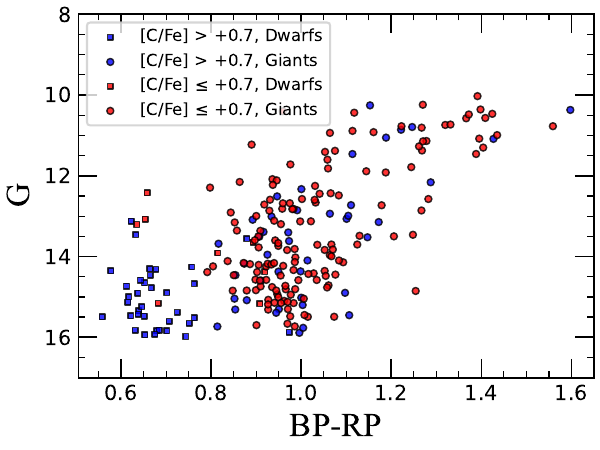}
\caption{Color-magnitude (observed color and apparent magnitude) diagram for sub-samples of the VMP/EMP stars shown in Figure \ref{fig:test-cfe}, represented by the red ([C/Fe] $\leq$ +0.7) and blue ([C/Fe] $>$ +0.7) symbols, respectively. The squares and circles represent dwarfs and giants, respectively.}
\label{fig:CMD_cfe}
\end{figure}

Figure \ref{fig:test-cfe} shows comparisons of the photometric-metallicity estimates we obtain with the spectroscopic determinations, color-coded by [C/Fe]. The $\rm [Fe/H]_{spec}$ and [C/Fe] values are originally from \citet{Yoon_2016}, \citet{Li_2022}, \citet{Placco_2022}, and \citet{Zepeda_2023}. Figure \ref{fig:CMD_cfe} is a color-magnitude diagram for the VMP/EMP samples of stars with [C/Fe] $\leq$ +0.7 and [C/Fe] $>$ +0.7, indicated by red and blue symbols, respectively. 

From inspection of Figure \ref{fig:test-cfe}, all three samples exhibit a good correlation between $\rm [Fe/H]_{GALEX}$ and $\rm [Fe/H]_{spec}$, with small offsets. Note that the dispersion value is larger for the sub-sample of stars with [C/Fe] $>$ +0.7 than for the sub-sample with [C/Fe] $<$ +0.7. This is likely due to 1) the fact  that the sub-sample of stars with [C/Fe] $>$ +0.7 are bluer (and less sensitive to [Fe/H]) and fainter (larger photometric errors), and thus have larger uncertainties in the $\rm[Fe/H]_{GALEX}$ estimates and, 2) the effects of carbon enhancement on our estimates, primarily for the cooler giants in our sample. However, we point out that, for all three samples, there are stars with photometric-metallicity estimates that straddle the one-to-one line, which is superior to the behavior seen by \citet{Hong2023} when comparing to photometric-metallicity estimates obtained using the u- and v-band filters from the SAGES and SMSS samples (see their Figure A1, where the carbon-enhanced stars often result in higher-metallicity estimates than their spectroscopic values). A similar problem is encountered in the analysis of \citet{Martin2023}, which used the Gaia XP spectra in combination with synthetic photometry of the Pristine survey CaHK magnitudes (see their Figure 17). As demonstrated by their comparison with the SAGA catalog of spectroscopic determinations \citep{saga2008}, the stars that are strongly carbon enhanced exhibit a clear bias, due to the influence of strong molecular carbon bands on the region near the Ca II H and K lines.
As a result, such stars have over-estimated Pristine metallicities, on the order ot 0.5 to 1.5\,dex, in particular at lower effective temperatures.

We conclude that the addition of the NUV-band can provide for improved estimates of photometric metallicity for stars with strong carbon enhancements. Clearly, having an independent estimate of [C/Fe], which is possible with the filter sets used by J-PLUS and S-PLUS — notably the J0430 filter, designed to detect prominent stellar absorption features of molecular carbon — would provide an even better alternative.

\section{The Final Sample} \label{sec:final}

In this section, we cross-match the entire Gaia EDR3 data and GALEX GR6+7 AIS data to create our (nearly) all-sky sample. 
After imposing the following selection criteria:
\begin{enumerate}
\item GALEX: error$_{\rm NUV} <$ 0.15 and NUV $>$ 15 to avoid saturation
\item Gaia (same as \citealt{Xu2022gaia-metal}, except that we extend the bright G magnitude limit to 6): sources with $|b|>10^\circ$, $\rm E(B-V)\leq0.5$, $\rm 6<G<16$, and limit on phot\_bp\_rp\_excess\_factor and BP$-$RP color, \end{enumerate}
4,468,105 dwarfs and 503,700 giants remain in our final sample (4,971,805 stars in all). Please note that the $|Z|>300$\,pc cut discussed in Section \ref{subsec:test-dr8} is not applied to our final sample. The color–magnitude diagram for this sample is shown in Figure \ref{fig:final-sample}.
The G-magnitude distributions of the final sample (with flag=0) are shown in 
Figure \ref{fig:G-distri}. 

We apply our model to this final sample. The [Fe/H] (with flag=0) distributions are shown in Figure \ref{fig:feh-distri}. The increase in the numbers of stars at the minimum [Fe/H] values is due to 1) the metallicities in our model can only be reliably determined in the range of $-4.0 <$ [Fe/H] $\leq +1.0$ for dwarfs and 
$-4.5 < $ [Fe/H] $\leq +1.0$ for giants, respectively, resulting in an accumulation at the lower boundaries of the range and, 2) there still remain stars with likely stellar activity, although we have applied a flag in an attempt to filter them out.
For dwarfs, there are 14,528 VMP and 2,972 EMP and stars, respectively. For giants, there are 18,747 VMP and 3,656 EMP stars, respectively.

The spatial distributions in the Z-R plane are shown in Figure \ref{fig:final-density}, where Z is the distance to the Galactic plane, and R is the Galactocentric distance. The spatial distributions, color-coded by [Fe/H], are shown in Figure  \ref{fig:feh-RZ}. The evident vertical gradients in metallicity within the disk-system stars are observable in both panels.
    
The columns contained in the final sample catalog are listed in Table \ref{table2}. Similar to \citet{Huang2023}, the radial-velocity values in the catalog are collected from five spectroscopic surveys: GALAH DR3+ (\citealt{Buder2021}), SDSS/ APOGEE DR17 (\citealt{Abdurro'uf2022}), Gaia DR3 (\citealt{Katz2023}), LAMOST DR9\footnote {\href{http://www.lamost.org/dr9/v1.0/}{http://www.lamost.org/dr9/v1.0/} (medium resolution catalog and low resolution catalog)
} and SDSS/SEGUE DR16 (\citealt{Ahumada2020}). In total, over 4,702,675 stars in our final sample have radial-velocity measurements. When a star has two or more radial-velocity measurements, we adopt the results from the highest spectral resolution survey. Note that we calibrated all of the radial-velocity offsets based on the APOGEE radial-velocity values. 
The catalog is publicly available at china-VO {\href{http://doi.org/10.12149/101365}{http://doi.org/10.12149/101365}.

\begin{figure}[htbp]
\centering
\includegraphics[width=8cm]{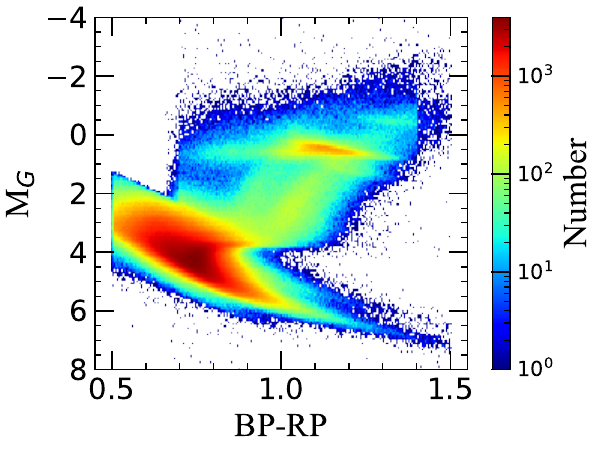}
\caption{Color-magnitude diagram of the final sample, color-coded by source number density.}
\label{fig:final-sample}
\end{figure}

\begin{figure*}[htbp]
\centering
\includegraphics[width=16cm]{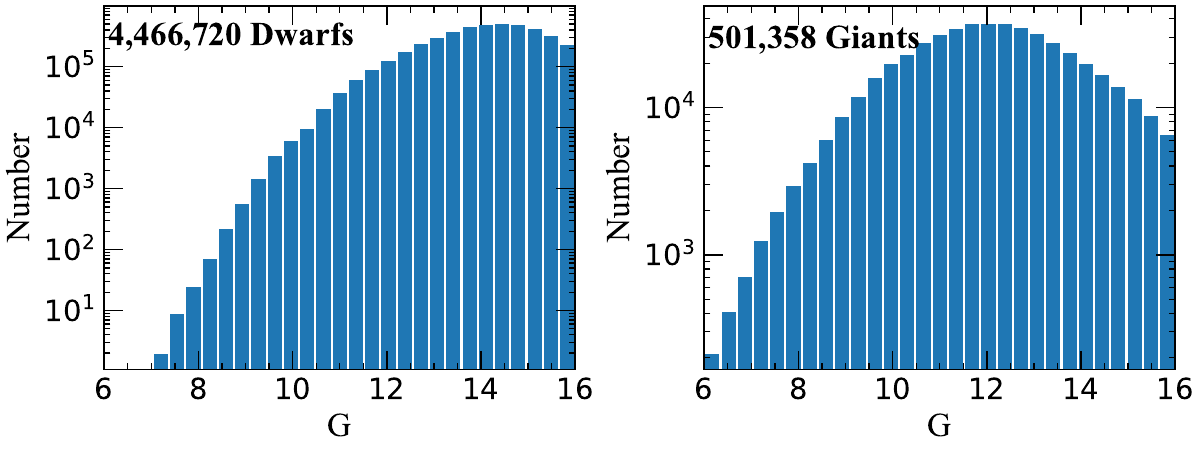}
\caption{The G-magnitude distribution of the final sample (with flag=0). The left and right panels are for the dwarfs and giants, respectively.}
\label{fig:G-distri}
\end{figure*}   

\begin{figure*}[htbp]
\centering
\includegraphics[width=16cm]{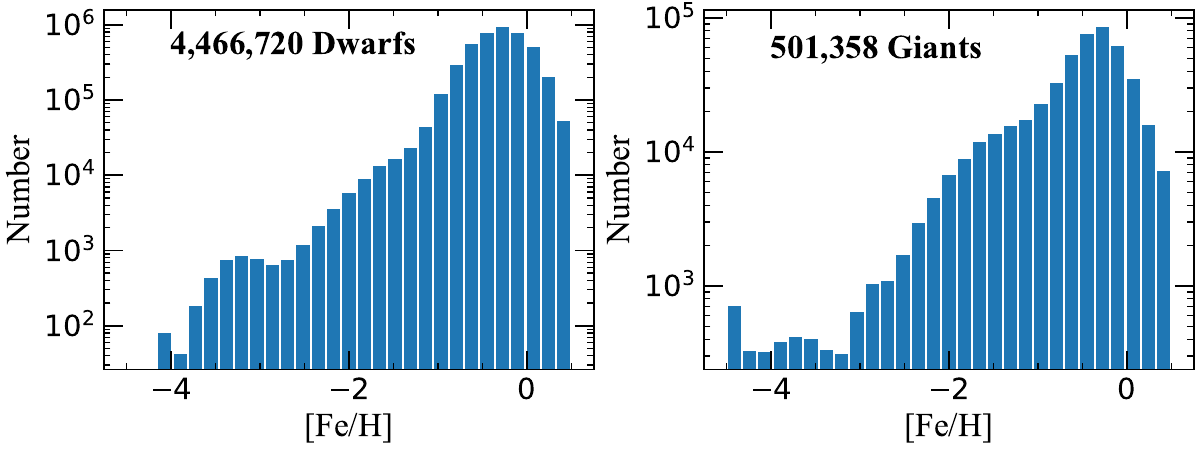}
\caption{The [Fe/H] distribution of the final sample (with flag=0). The left and right panels are for the dwarfs and giants, respectively. For dwarfs, there are 14,522 VMP and 2,962 EMP stars, respectively.  For giants, there are 18,245 VMP and 3,582 EMP stars, respectively. Note that the increase in the numbers of stars at the lowest metallicities are due to limitations of our model and the likely presence of stellar activity (see text).}
\label{fig:feh-distri}
\end{figure*}   

\begin{figure*}[htbp]
\centering
\includegraphics[width=16cm]{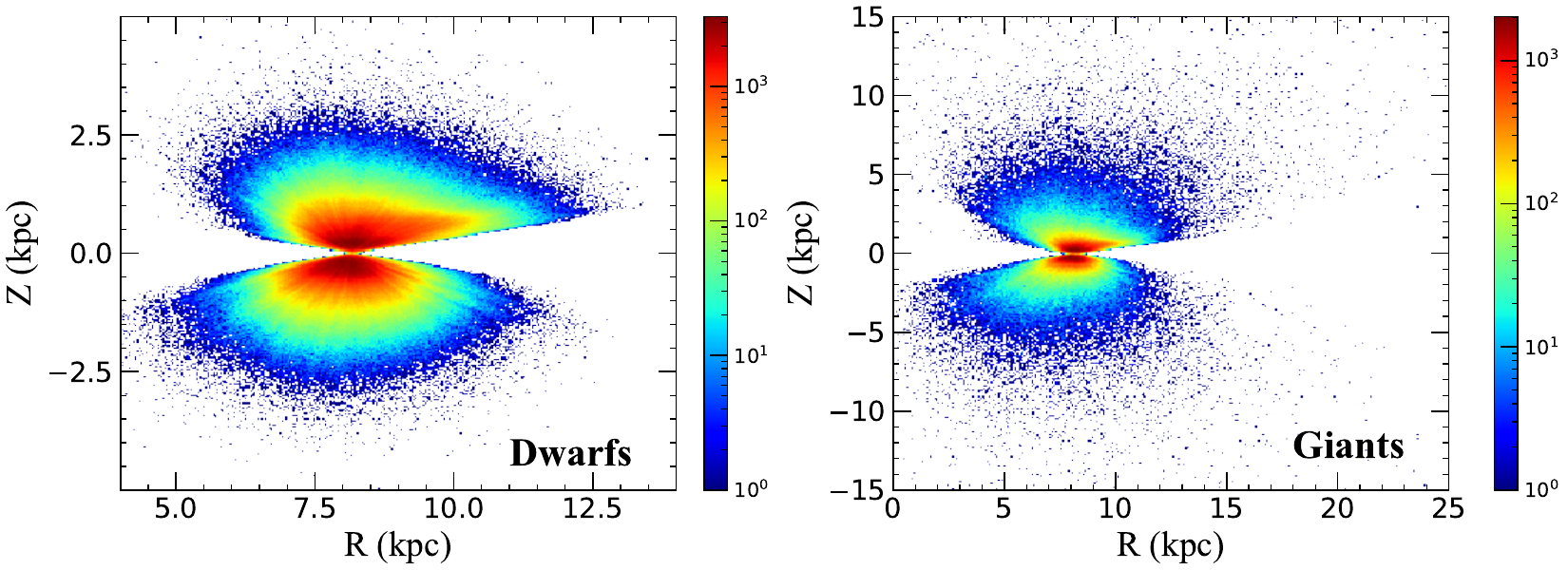}
\caption{The spatial distributions in the Z--R plane of the final sample, color-coded by number densities. The left and right panels are for the dwarfs and giants, respectively. The Sun is located at (R, Z) = (8.178, 0.0) kpc. }
\label{fig:final-density}
\end{figure*}   

\begin{figure*}[htbp]
\centering
\includegraphics[width=16cm]{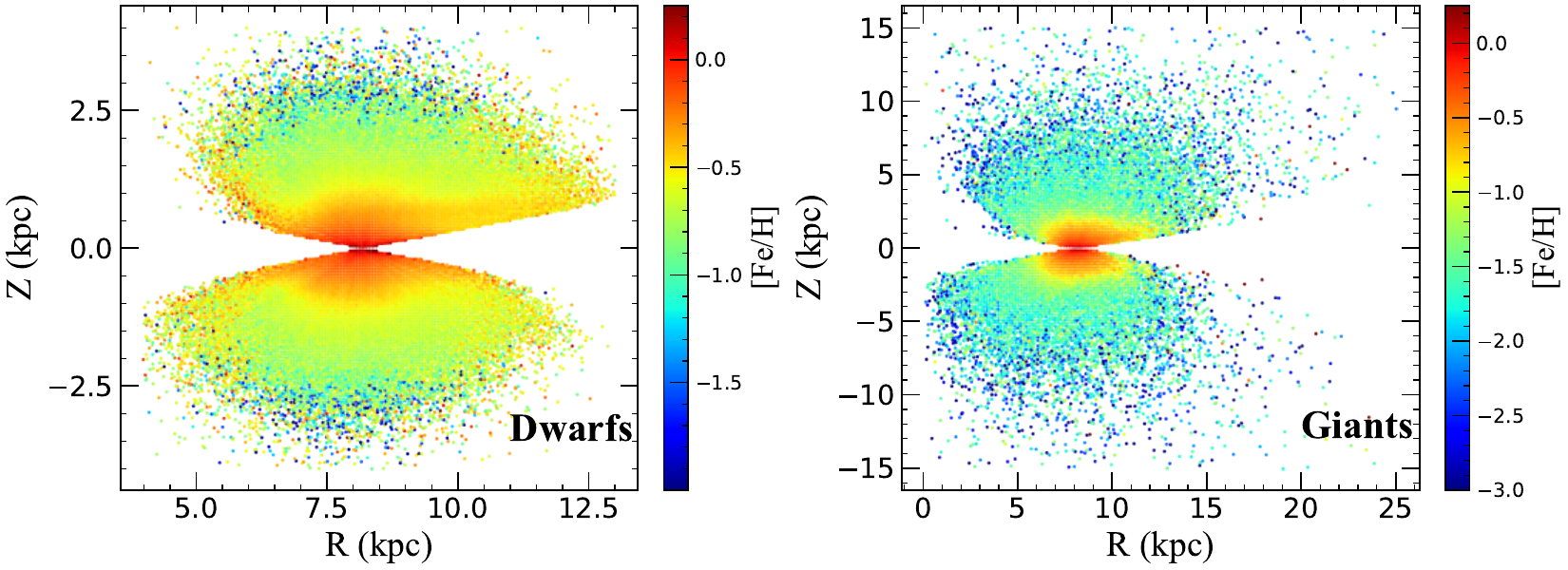}
\caption{The spatial distributions in the Z--R plane of the final sample, color-coded by $[\rm Fe/H]_{GALEX}$. The left and right panels are for the dwarfs and giants, respectively. The Sun is located at (R, Z) = (8.178, 0.0) kpc. }
\label{fig:feh-RZ}
\end{figure*}

\setlength{\tabcolsep}{1.5mm}{
\begin{table*}[htbp]
\footnotesize
\centering
\caption{Description of the Final Sample}
\begin{tabular}{lll}
\hline
\hline
Field & Description & Unit\\
\hline
source\_id & Unique source identifier for Gaia EDR3 (unique with a particular Data Release) & --\\
ra & Right ascension from Gaia EDR3 & deg\\
dec & Declination from Gaia EDR3 & deg\\
parallax & Parallax from Gaia EDR3 & mas\\
parallax\_error & Standard error of parallax from Gaia EDR3 & mas\\
parallax\_corrected & Parallax corrected by \citet{Lindegren2021}& mas\\
pmra & Proper motion in right ascension direction from Gaia EDR3 & mas/year\\
pmra\_error & Standard error of proper motion in right ascension direction from Gaia EDR3 & mas/year\\
pmdec & Proper motion in declination direction from Gaia EDR3 & mas/year\\
pmdec\_error & Standard error of proper motion in declination direction from Gaia EDR3 & mas/year\\
ruwe & Renormalised unit weight error from Gaia EDR3 & --\\
phot\_bp\_rp\_excess\_factor & BP/RP excess factor from Gaia EDR3 & --\\
phot\_bp\_mean\_mag & Integrated BP-band mean magnitude from Gaia EDR3 & --\\
phot\_rp\_mean\_mag & Integrated RP-band mean magnitude from Gaia EDR3 & --\\
phot\_g\_mean\_mag & G-band mean magnitude from Gaia EDR3 & --\\
l & Galactic longitude from Gaia EDR3 & deg\\
b & Galactic latitude from Gaia EDR3 & deg\\
feh\_gaia & Photometric metallicity from \cite{Xu2022gaia-metal} & -- \\
feh\_Gaia\_error & Formal error of feh\_gaia from \cite{Xu2022gaia-metal}  & dex\\
ebv & Value of $\rm E(B-V)_{SFD}$ & --\\
type & Flag to classify stars (\citealt{Xu2022gaia-metal});  0 for dwarfs and 1 for giants & --\\
NUVmag & NUV calibrated magnitude & --\\
e\_NUVmag & NUV calibrated magnitude error & mag\\
BP\_RP0 & Intrinsic BP$-$RP color after color correction & --\\
NUV\_BP0 & Intrinsic NUV$-$BP color after color correction & --\\
feh\_GALEX\_old & Photometric metallicity derived from previous model without consideration of $\rm M_G$  & --\\
feh\_GALEX & Photometric metallicity derived from New Model with consideration of $\rm M_G$  & --\\
M\_G & Absolute Magnitude of G-band, calculated by: \\
 & $\rm M_{G}= G - 10 - 5 \times  \log_{10}{\frac{1}{parallax\_corrected}} -R_{G}\times E(B-V) $& --\\
flag$^1$ & The quality flag & --\\
rpgeo & Median of the photogeometric distance posterior (r\_med\_photogeo) from \citet{Bailer-Jones2021} & pc\\
b\_rpgeo & 16th percentile of the photogeometric distance posterior (r\_lo\_photogeo) from \citet{Bailer-Jones2021} & pc\\
B\_rpgeo & 84th percentile of the photogeometric distance posterior (r\_lo\_photogeo) from \citet{Bailer-Jones2021} & pc\\
rv & Radial velocity & km/s\\
e\_rv & Error of radial velocity & km/s\\
flag\_rv & Flag to indicate the source of radial velocity, which takes the values “GALAH,” “APOGEE,” “Gaia,”\\ &“LAMOST\_MRS,” “LAMOST\_LRS,” “SEGUE” & --\\
\hline
\end{tabular}
\label{table2}

\tablecomments{
The flag takes the values of 0 or 1.
`flag=1' means the stars outside the limits where we consider the feh\_GALEX values to be reliable. `flag=0' means the stars are within the limits, and considered reliable.}
\end{table*}}

\section{Summary} \label{sec:summary}
In this work, we combine the SAGA and PASTEL catalogs with Gaia EDR3 and GALEX GR6+7 AIS data as our training sample to construct a relational model from stellar loci that incorporate $\rm M_G$ to estimate photometric metallicitities for dwarfs and giants. We construct an all-sky sample from Gaia EDR3 and GALEX GR6+7 AIS data as our final sample (4,468,105 dwarfs and 503,700 giants), employing a precise reddening correction using empirical temperature- and extinction-dependent extinction coefficients. 

We first obtain the [Fe/H]- and $\rm M_G$-dependent stellar loci for the NUV$-$BP and BP$-$RP colors. For a given $\rm BP-RP$ color, a 1\,dex change in [Fe/H] causes an approximate 1 mag change in $\rm NUV-BP$ color for Solar-type stars. The [Fe/H]- and $\rm M_G$-dependent stellar loci are then used to estimate stellar photometric metallicities from $\rm NUV-BP$,  BP$-$RP, and $\rm M_G$. As a result of the very strong metallicity dependence in the NUV-band and the precise reddening corrections, we have obtained a typical precision of 0.11\,dex for dwarfs and 0.17\,dex for giants, with an effective metallicity range from $-3.0 <$ [Fe/H] $\leq +0.5$ for dwarfs and $-4.0 <$ [Fe/H] $\leq +0.5$ for giants. 
We also demonstrate that the NUV-band based photometric-metallicity estimate is not as strongly affected by carbon enhancement as previous photometric techniques. Moreover, our tests confirm that incorporating absolute magnitude ($\rm M_G$) in the stellar loci serves to further improve the accuracy of the photometric-metallicity estimates. 

Our work demonstrates the feasibility and great potential of photometric-metallicity estimates using the NUV-band photometry, due to its very strong sensitivity to [Fe/H]. A number of lessons have been learned in this exploration, including:
\begin{enumerate}
\item The NUV-band photometry also exhibits a strong dependence on surface gravity, therefore, the absolute magnitude ($\rm M_G$ in this work) has to be taken into account in the modeling
\item The reddening correction in the NUV-band is essential and challenging. In addition to the effect of surface gravity, even the effect of metallicity on the extinction coefficient needs to be considered
\item Stellar activity in cool dwarf and giant stars could cause a significant under-estimation of metallicity, thus should be considered when hunting for VMP/EMP stars\item Future improvement in the calibration precision of the NUV-band photometry is very important
\item With future improvement in the calibration precision in the NUV-band photometry, it may be possible to identify likely carbon-enhanced VMP/EMP stars by a comparison of 
photometric-metallicity estimates based on the NUV-band with other bands, such as the u- and 
v-bands. 
\end{enumerate}

Our final sample, encompassing a wide range of [Fe/H] values, is expected to be valuable for numerous research studies. For instance, it can provide candidate metal-poor stars for subsequent high-resolution spectroscopic follow-up, and also for chemo-dynamical studies of MW stellar populations. 
Compared with other massive, all-sky catalogs of metallicities, the metallicities in this work are measured based on the NUV-band data, which are measurements utilizing independent information. Therefore, our catalog is complementary to other catalogs. In addition, given that the NUV-band is hardly influenced by C- and N-enhancement, our measurements could be used to provide a more complete census of VMP and EMP stars, and to select C-enhanced VMP stars and N-enhanced stars (e.g., \citealt{Zhang_jiajun_2023}) when combined with other datasets. Such explorations will be presented in future work.
This work lays the foundation for using NUV-band data from space telescopes such as the upcoming Chinese Space Station Telescope (CSST; \citealt{CSST_2011}).
\clearpage
\begin{acknowledgments}
The authors thank the referee for his/her suggestions that improved the clarity of our presentation.
This work is supported by the National Natural Science Foundation of China through the project NSFC 12222301, 12173007 and 11603002,
the National Key Basic R\&D Program of China via 2019YFA0405503 and Beijing Normal University grant No. 310232102. 
We acknowledge the science research grants from the China Manned Space Project with NO. CMS-CSST-2021-A08 and CMS-CSST-2021-A09. T.C.B. and J.H. acknowledge partial support from grant PHY 14-30152,
Physics Frontier Center/JINA Center for the Evolution
of the Elements (JINA-CEE), and from OISE-1927130: The International Research  Network for Nuclear Astrophysics (IReNA), awarded by the US National Science Foundation. T.C.B.'s participation in this work was initiated by conversations that took place during a visit to China in 2019, supported by a PIFI Distinguished Scientist award from the Chinese Academy of Science.

This work has made use of data from the European Space Agency (ESA) mission Gaia (\url{https://www.cosmos.esa.int/gaia}), processed by the Gaia Data Processing and Analysis Consortium (DPAC, \url{https:// www.cosmos.esa.int/web/gaia/dpac/ consortium}). Funding for the DPAC has been provided by national institutions, in particular, the institutions participating in the Gaia Multilateral Agreement. 
Guoshoujing Telescope (the Large Sky Area Multi-Object Fiber Spectroscopic Telescope LAMOST) is a National Major Scientific Project built by the Chinese Academy of Sciences. Funding for the project has been provided by the National Development and Reform Commission. LAMOST is operated and managed by the National Astronomical Observatories, Chinese Academy of Sciences.
We acknowledge the invaluable contribution of GALEX (NASA's Galaxy Evolution Explorer) in providing the dataset used in this research.
\end{acknowledgments}

\bibliographystyle{aasjournal}
\bibliography{GALEX_metal}

\begin{appendix}

\section{Motivation and Impact of Absolute Magnitude}
\label{sec:appendix}
In our initial work, we used the relationship between stellar loci (NUV$-$BP and BP$-$RP colors) and metallicity to establish a model for estimation of photometric metallicities, as shown in Figure \ref{fig:old-model}. 
The residuals $\Delta [\rm Fe/H]_{old}$ in Figure \ref{fig:motivation} are calculated from this initial model as $\rm [Fe/H]_{GALEX_{old}} - [Fe/H]_{HR}$. Moderate correlation between $\Delta [\rm Fe/H]_{old}$ and $\rm M_G$ has been found in our training and test samples. This result is not surprising, due to the fact that surface gravity also has a strong effect on the $BP-NUV$ color. Considering that sensitivity to $\rm M_G$ is different in different color ranges, we have selected sources from the training sample with $\rm 0.95<BP-RP<1.15$ for dwarfs and $\rm 0.85<BP-RP<1$ for giants, to demonstrate the influence of $\rm M_G$ on photometric metallicities more clearly. The results are shown in Figure \ref{fig:motivation}. The fact that $\rm M_G$ affects our photometric metallicities suggests that incorporating $\rm M_G$ is essential. Consequently, we refine our model by factoring in the impact of $\rm M_G$, and indeed obtain better results than before.

The $\Delta [\rm Fe/H]_{old}$ in the middle column of panels of Figure \ref{fig:diff-M_G} is also calculated from our initial model as $\rm [Fe/H]_{GALEX_{old}} - [Fe/H]_{LAMOST}$. Figure \ref{fig:diff-M_G}, which shows the comparison with LAMOST DR8, is an effective illustration of the difference (depicted by $\Delta [\rm Fe/H]$) between the photometric metallicities and LAMOST metallicities. 
In the middle column of panels in Figure \ref{fig:diff-M_G}, we observe a distinct pattern in the distribution of $\Delta [\rm Fe/H]_{old}$ with respect to $\rm M_G$. As $\rm M_G$ increases, $\Delta [\rm Fe/H]_{old}$ exhibits a structured variation, especially among metal-poor giants. Such variations are particularly noticeable in the area within the black rectangular box, with the trends being consistent with those in Fig\,\ref{fig:motivation}.

Based on these observations, we decided to incorporate absolute magnitude ($\rm M_G$) into our model. The residuals $\Delta [\rm Fe/H]_{new}$ in the right column of panels of Figure \ref{fig:diff-M_G} is calculated from our new model as $\rm [Fe/H]_{GALEX_{new}} -[Fe/H]_{LAMOST}$. This diagram demonstrates the changes in $\Delta [\rm Fe/H]_{new}$ on the color-magnitude diagram, with black contours representing star density. The $\Delta [\rm Fe/H]$ values of the majority of stars are close to 0.0\,dex, indicating that our model is well-suited for the majority of giants. Furthermore, it is evident that the previously observed variation pattern is much weaker after considering $\rm M_G$ in our new model. 
Additionally, the typical deviations from $\Delta[\rm Fe/H]=0$ have been reduced, as indicated by the legends in each panel. 
However, there are a minority of $\Delta[\rm Fe/H]$ values that display deviations from zero  among the bluer $\rm BP-RP$ colors, shown in the shaded region in the bottom-middle panel. These stars could be interpreted as
blue horizontal-branch stars and stars situated near the boundary between dwarfs and giants. Their properties slightly differ from those of most FGK giants in our sample. In any case, as illustrated here, this new version performs better compared with the old test result, successfully resolving the previously discernible trends related to $\rm M_G$.

\newcounter{Afigure}
\setcounter{Afigure}{1}
\renewcommand{\thefigure}{A\arabic{Afigure}}

\begin{figure*}[htbp]
\centering
\subfigure{
\includegraphics[width=7cm]{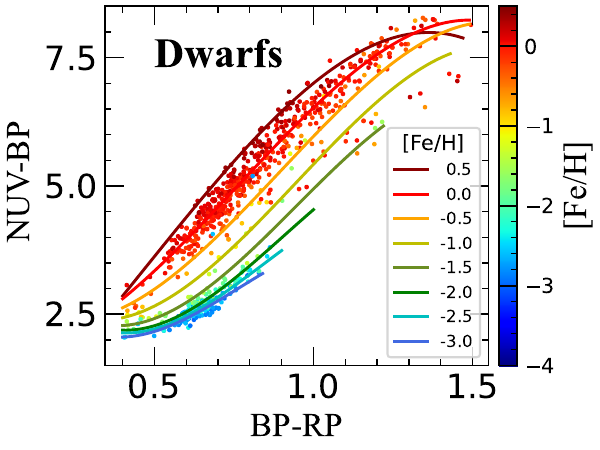}
}
\subfigure{
\includegraphics[width=7cm]{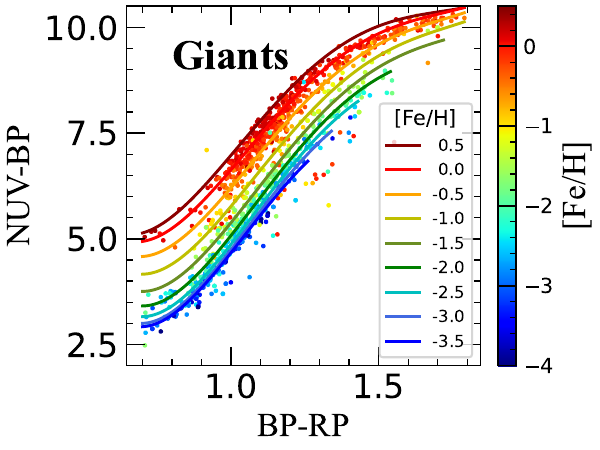}
}
\caption{Distribution of the training sample in the ($\rm NUV-BP$) vs. ($\rm BP-RP$) plane, color-coded by [Fe/H], with dwarfs in the left panel and giants in the right panel. The colored lines represent different [Fe/H] values, ranging from  to $-$3.5 to $+$0.5, in steps of 0.5\,dex.
}
\label{fig:old-model}
\end{figure*}

\stepcounter{Afigure}
\begin{figure*}[htbp]
\centering
\subfigure{
\includegraphics[width=7cm]{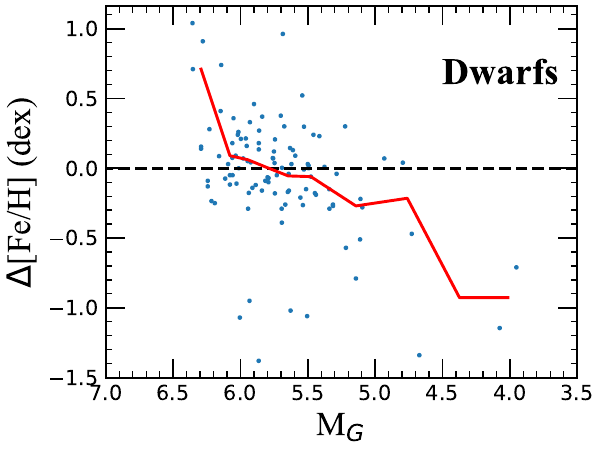}
}
\subfigure{
\includegraphics[width=7cm]{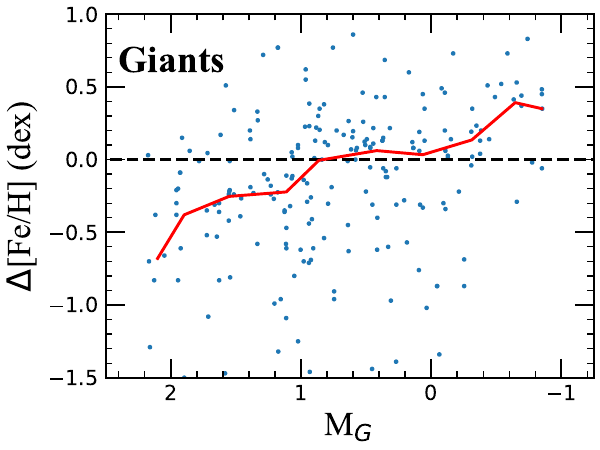}
}

\caption{Distribution of the $\Delta[\rm Fe/H]$ (calculated from the model without $\rm M_G$) as a function of $\rm M_G$ for dwarfs of $\rm 0.95<BP-RP<1.15$ (left panel) and giants of  $\rm 0.85<BP-RP<1$ (right panel)}, respectively. The 
red-solid line indicates the median values.

\label{fig:motivation}
\end{figure*}

\stepcounter{Afigure}
\begin{figure*}[htbp]
\centering
\includegraphics[width=16cm]{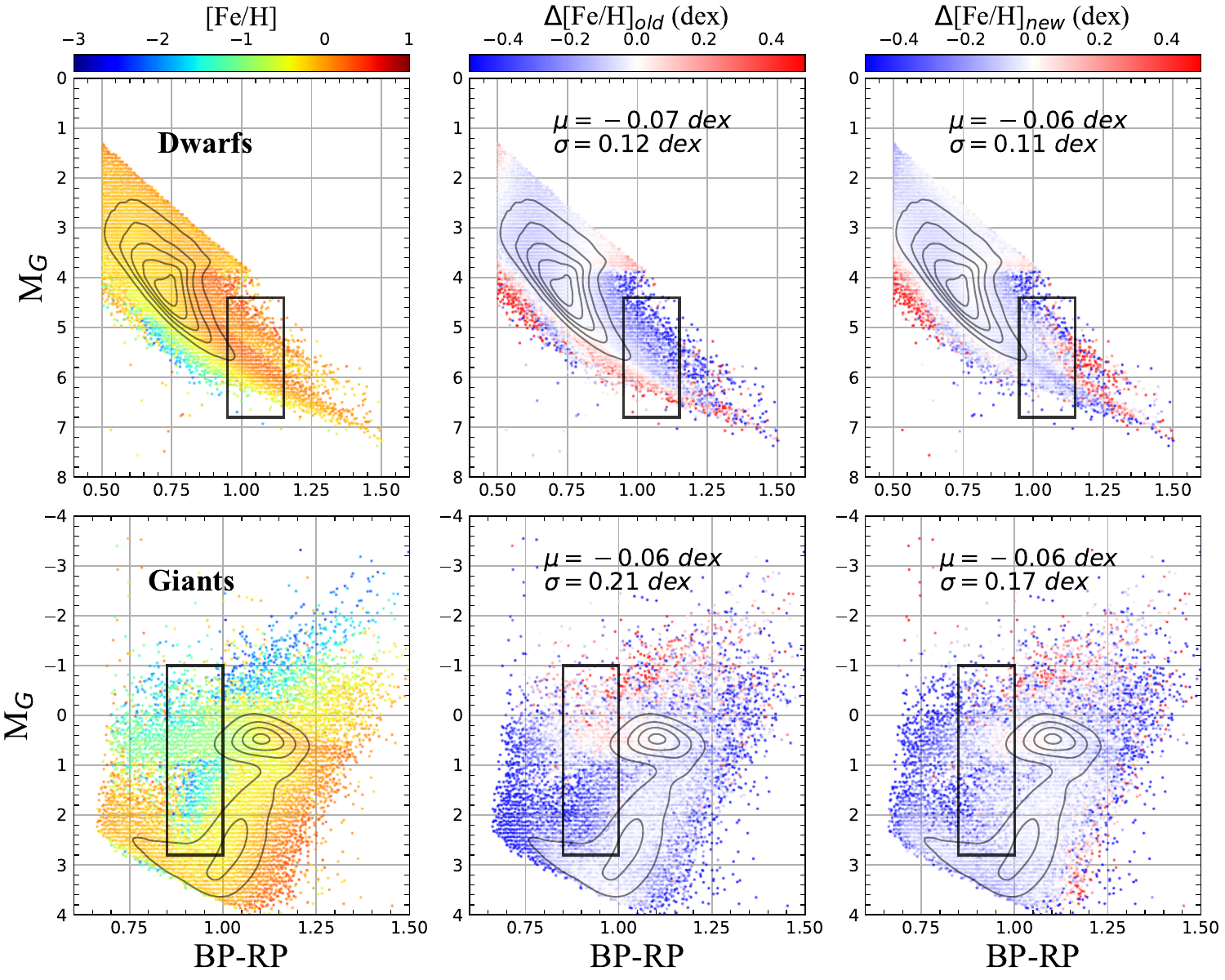}
\caption{Color-magnitude diagram of the LAMOST DR8 test sample. Dwarfs are shown in the upper row of panels and giants are shown in the lower row of panels. The plots are color-coded by $[\rm Fe/H]_{LAMOST}$ (left column of panels), $\Delta[\rm Fe/H]$ (old model without $\rm M_G$) (middle column of panels), and $\Delta[\rm Fe/H]$ (new model with $\rm M_G$ considered) (right column of panels). In the lower middle panel, the shaded region indicates possible blue horizontal-branch stars or stars situated near the boundary between dwarfs and giants.
}
\label{fig:diff-M_G}
\end{figure*} 

\end{appendix}

\end{document}